\documentclass[a4paper,11pt]{article}
\usepackage{aaskaiid}
\usepackage{orcidlink}
\usepackage{hyperref}
\usepackage{comment}
\title{The Astrophysics of Fast Radio Bursts}
\ShortTitle{Astrophysics of FRBs}

\author[1,2,3]{Alice P. Curtin\orcidlink{0000-0002-8376-1563}}
\author[4]{Marcin Gawro\'nski\orcidlink{0000-0003-4056-4903}}
\author[1,2,3,5]{Jason Hessels\orcidlink{0000-0003-2317-1446}}
\author[6]{Clancy James\orcidlink{0000-0002-6437-6176}}
\author[7]{Fabian Jankowski\orcidlink{0000-0002-6658-2811}}
\author[5,9]{Franz Kirsten\orcidlink{0000-0001-6664-8668}}
\author[5, 9]{Benito Marcote\orcidlink{0000-0001-9814-2354}}
\author[10]{Harry Qiu\orcidlink{0000-0002-9586-7904}}
\author[11]{Robert Reischke\orcidlink{0000-0001-5404-8753}}
\author[1,2]{Mawson W. Sammons\orcidlink{0000-0002-4623-5329}}
\author[12]{Laura G. Spitler\orcidlink{0000-0002-3775-8291}}
\author[13]{Ben Stappers\orcidlink{0000-0001-9242-7041}}
\author[14, 15]{Amanda Weltman\orcidlink{0000-0002-5974-4114}}
\author[]{The SKA Transients SWG}
\ShortName{Coordinator: Alice P. Curtin} % shortened name list for header

\affiliation[1]{Department of Physics, McGill University, 3600 rue University, Montr\'eal, QC H3A 2T8, Canada}
\affiliation[2]{Trottier Space Institute, McGill University, 3550 rue University, Montr\'eal, QC H3A 2A7, Canada}
\affiliation[3]{Anton Pannekoek Institute for Astronomy, University of Amsterdam, Science Park 904, 1098 XH Amsterdam, The Netherlands}
\affiliation[4]{Institute of Astronomy, Faculty of Physics, Astronomy and Informatics, Nicolaus Copernicus University, Grudziadzka 5, PL-87-100, Toru\'n, Poland}
\affiliation[5]{ASTRON, Netherlands Institute for Radio Astronomy, Oude Hoogeveensedijk 4, 7991 PD Dwingeloo, The Netherlands}
\affiliation[6]{International Centre for Radio Astronomy Research, Curtin University, Bentley, 6102, WA, Australia}
\affiliation[7]{LPC2E, OSUC, University of Orleans, CNRS, CNES, Observatoire de Paris, F-45071 Orleans, France}
\affiliation[8]{Department of Space, Earth and Environment, Chalmers University of Technology, Onsala Space Observatory, 439 92, Onsala, Sweden}
\affiliation[9]{Joint Institute for VLBI ERIC, Oude Hoogeveensedijk 4, 7991PD Dwingeloo, The Netherlands}
\affiliation[10]{SKA Observatory, Jodrell Bank, Lower Withington, Macclesfield SK11 9FT, UK}
\affiliation[11]{Argelander-Institut für Astronomie, Universität Bonn, Auf dem Hügel 71, D-53121 Bonn, Germany}
\affiliation[12]{Max-Planck-Institut für Radioastronomie, Auf dem Hügel 69, 53121 Bonn, Germany}
\affiliation[13]{Jodrell Bank Centre for Astrophysics, Department of Physics and Astronomy, The University of Manchester, Manchester M13 9PL, UK}
\affiliation[14]{High Energy Physics, Cosmology and Astrophysics Theory (HEPCAT) Group, Department of Mathematics and Applied Mathematics, University of Cape Town, Rondebosch, Cape Town, 7700, South Africa}
\affiliation[15]{African Institute for Mathematical Sciences, 6 Melrose Road, Muizenberg, Cape Town, 7945, South Africa}

\emailAdd{alice.curtin@mail.mcgill.ca}

\abstract{Fast radio bursts (FRBs) provide a  glimpse of high-energy astrophysical phenomena in other galaxies. They point the way to extreme conditions that are currently undetectable by any other known means. These coherent radio flashes have timescales of microseconds to milliseconds, and inferred energies that are comparable to those of the most extreme bursts seen from Galactic neutron stars. However, the nature of FRB sources remains an open question in astrophysics. Magnetically powered neutron stars known as `magnetars' are a leading candidate for explaining the FRB phenomenon, but other plausible progenitors include magnetically interacting neutron-star binaries or accreting black holes. The diversity of FRB burst types and their galactic environments hint that multiple mechanisms and progenitor types may be responsible. Here we discuss the ways in which the SKA can uncover the nature of FRBs. In particular, we focus on the key advantages of the SKA: its Southern Hemisphere location and hence overlapping sky coverage with the Vera C. Rubin Observatory, its high sensitivity compared to existing wide-field FRB surveys, its fast search timescales down to tens of $\mu$s, and its broad spectral coverage with bands from 50 MHz to 15 GHz. With these capabilities, the SKA will excel in detecting FRB sources across new frequency ranges and timescales. This will aid in a better understanding of the fundamental astrophysics behind FRBs, which will in turn also contribute to their use as cosmological probes, as explored in companion chapters.}

\begin{document}
\maketitle

\newpage
\section{Introduction}

The discovery of pulsars in 1967 by Jocelyn Bell Burnell and collaborators revealed the existence of neutron stars and provided a powerful observational tool for studying the physics of such compact objects \citep{JocelynBelllDiscoveryNature}. Forty years later, the discovery of the first fast radio burst (FRB) by \citet{LorimerBurst} established an even more extreme type of impulsive radio emission, providing another laboratory for studying the extremes of the Universe.

Observationally, FRBs can be defined as roughly millisecond-duration radio flashes whose dispersion measures (DMs) imply an extragalactic origin. Broadly, there are at least two classes of FRBs: those that sporadically repeat and those that are apparent one-off events. In addition to their activity rates, systematic differences in the emission bandwidth and timescale of repeaters vs.\ non-repeaters also suggest that they derive from distinct mechanisms and/or progenitors \citep{pgk+21, 2025ApJ...992..206C}.

It took over a decade to firmly establish the astrophysical origins of FRBs (as opposed to artificial interference), to collect the first small sample of signals, and to localise a few to their host galaxy \citep{Thornton_2013, 2014ApJ...790..101S, Chatterjee2017}. In the past five years, however, progress has rapidly accelerated. Thousands of FRB sources are now known \citep{SecondCHIMEFRBCatalog}, over a hundred have been localised to their host galaxy (e.g.,~\citealt{2020ApJ...895L..37B, 2024ApJ...967...29L, 2025arXiv250705982P, 2025ApJS..280....6C}), and monitoring of repeating sources has revealed their dynamic local environments \citep[e.g., ][]{michilli2018extreme, 2022Natur.609..685X, 2023ApJ...950...12M, 2023Sci...380..599A, 2025arXiv250916374O, Pandhi_2026}.

The short timescales ($\mu$s - ms) and enormous luminosities ($\sim10^{36}-10^{44}$  erg s$^{-1}$) of FRB emission necessitate a compact and high-energy-density environment. Additionally, to produce such high luminosities on such short timescales, a coherent emission mechanism is required. Theoretical models for the nature of FRBs have thus focused on white dwarfs (WDs), neutron stars (NSs), and black holes (BHs) with a variety of emission mechanism models \citep[e.g., ][]{klb17, lu2018radiation, mms19}. However, the exact underlying emission mechanism, progenitor type, and energy source are still not definitively confirmed.

\begin{figure}
    \centering
    \includegraphics[width=\linewidth]{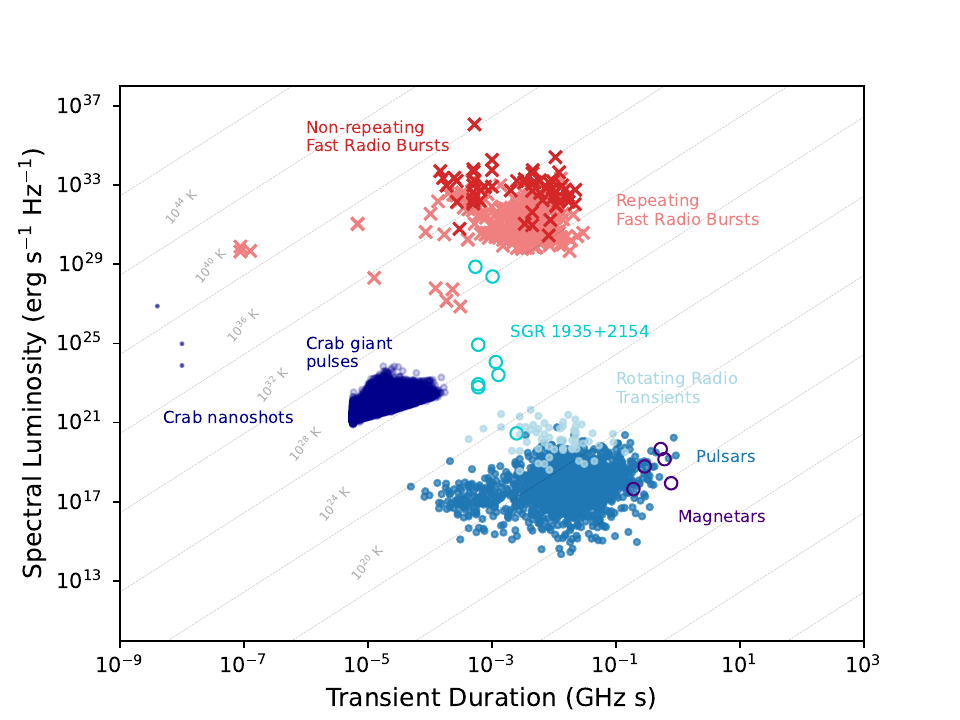}
    \caption{Spectral Luminosity vs. frequency-weighted transient duration for a sub-sample of pulsars, rotating radio transients (RRATs), Crab giant radio pulses (GRPs), Crab nanoshots, SGR 1935+2154 bursts, magnetar pulses, and FRBs. Figure adapted from that of \citet{2022NatAs...6..393N} with the additional inclusion of results from \citet{2023Sci...382..294R, 2024ApJ...967...29L, 2025ApJS..280....6C}. The FRBs span over a full 10 orders of magnitude in their luminosities, with the brightest bursts from the Galactic magnetar SGR 1935+2154 consistent in luminosity with the faintest repeating FRBs detected. Burst luminosities from SGR 1935+2154 similarly span a large seven to eight orders of magnitude in range \citep{2021NatAs...5..414K}. While not demonstrated here due to showing the frequency-weighted transient duration, non-repeating FRBs tend to have shorter temporal durations but larger bandwidths than bursts from repeating sources \citep{pgk+21, 2025ApJ...992..206C}.}
    \label{fig:TransientPhaseSpace}
\end{figure}

The discovery of an FRB-like burst from the Galactic magnetar SGR~1935+2154 suggests that at least some extragalactic FRBs arise from magnetars \citep{brb+20, abb+20}. However, the burst from SGR 1935+2154 was a few orders of magnitude less energetic than FRBs seen at extragalactic distances \citep[see Figure \ref{fig:TransientPhaseSpace}; ][]{2022NatAs...6..393N} and FRBs have been seen from a range of local host galaxy environments with a diverse set of burst properties \citep[e.g.,][]{Chatterjee2017,bkm+21, 2022Natur.602..585K, 2023ApJ...954...80G, ssl+24, edf+24}. Thus, it remains unclear whether magnetars can explain all FRB sources. As with short- and long-gamma-ray bursts (GRBs), the observed population of FRB signals may have multiple physical origins \citep{2014Berger, 2019GalYam}. 

Astronomers have only recently begun exploring the full parameter space for FRB discovery. While Figure  \ref{fig:TransientPhaseSpace} compares FRB luminosities and durations to certain Galactic events, it does not emphasize the different activity rates, polarization states, and spectral extents seen from FRBs. Future efforts to expand transient searches to a wider range of timescales, frequencies, and redshifts could be the key to revealing multiple astrophysical origins for FRBs. 

In this chapter, we outline how the SKA can contribute to understanding the origin(s) and astrophysics of FRBs. Two accompanying chapters titled `Probing the baryon distribution with Fast Radio Bursts' and `Fast Radio Bursts as Cosmological Probes' focus on how we can use FRB signals as tools to study other topics such as the density, clumpiness, and magnetisation of the intergalactic medium (IGM) as well as the larger-scale evolution of the Universe \citep{Caleb01.2026.SKA, Caleb02.2026.SKA}. Here we focus on the following interconnected, high-level questions:

\begin{itemize}
    \item Do all FRBs have the same astrophysical origin?
    \item Which astrophysical sources produce FRBs, and under what circumstances?
    \item Do all FRBs repeat, or are some related to cataclysmic explosions?
    \item What powers FRB emission and what is the mechanism?
    \item What can we learn about the astrophysics of WDs, NSs, and BHs using FRBs? 
    \item How do FRBs relate to the significantly less luminous radio transients we detect from Galactic sources?
    \item What physical properties of FRBs are intrinsic rather than due to propagation effects? How does this impact FRBs as cosmological probes?   
\end{itemize}

\section{How to reveal the nature and astrophysics of FRBs}

There is a rich literature describing possible FRB progenitors and emission mechanisms. We refer here to the progenitor as the astrophysical object and conditions that lead to FRB production while the emission mechanism is a physical process by which a progenitor emits an FRB. The high volumetric event rate of FRBs \citep{2019NatAs...3..928R} as well as their repeating nature \citep{2016Natur.531..202S} tells us that most FRB progenitors must be relatively long-lived sources of energy, though some small fraction of FRBs could come from cataclysmic events. \citet{2019PhR...821....1P} provide a catalogue of FRB theories, and describe progenitors like isolated magnetars \citep{2014MNRAS.442L...9L, 2020ApJ...896..142B}, accreting BHs \citep{2021ApJ...917...13S}, and compact binary WDs \citep{2014A&A...569A..86M}. Potential cataclysmic events include, e.g., binary NS, BH, or WD mergers \citep{2021ApJ...917L..11K, Cooper2023}. More exotic proposals involve cosmic strings \citep{2017arXiv170702397B}, axion interactions in NSs \citep{2015PhRvD..91b3008I, 2019PhRvD..99d3535V}, and BH `batteries' \citep{Mingarelli2015}.  \citet{2019PhR...821....1P} also describe the possible emission mechanisms by which FRBs are generated, which could be common to several progenitor types. These mechanisms include, e.g., coherent curvature radiation \citep[][]{klb17, lu2018radiation}, synchrotron masers \citep[][]{mms19}, and Dicke’s super-radiance \citep{2018MNRAS.475..514H}. 

Depending on their exact origins, FRBs could provide new insights into the underlying physics of WDs, NSs, and (stellar mass) BHs, all objects that are notoriously hard to detect, especially at extragalactic distances. Understanding the intrinsic properties and nature of FRBs will thus have direct implications for our understanding of high-energy astrophysics, compact objects, and possibly even dark matter, cosmic strings, or axion physics.

As with many open topics in astronomy, several complementary strategies are needed to further extend our understanding of FRBs: (1) a larger population of known sources to identify groupings and outliers, (2) deeper characterisation of individual sources, and (3) exploration of a wider observational parameter space. Below, we outline some of the key FRB observables that are relevant for these strategies. 

\textbf{FRB signal properties}: The spectral, temporal, and polarimetric properties of FRBs give hints to their emission mechanisms and serve as a way to quantitatively compare different sources \citep{2019ApJ...876L..23H, 2021Natur.598..267L, 2025Natur.637...43M, 2025Natur.637...48N}. The observed signals are also influenced by propagation effects incurred in the intervening magneto-ionic media along the line-of-sight, giving us direct insights into both the local environments of FRBs \citep{michilli2018extreme, 2022Sci...375.1266F, 2023Sci...380..599A} as well as the intervening media \citep[see accompanying chapters on FRBs as cosmological probes; ][]{Caleb01.2026.SKA, Caleb02.2026.SKA}. These effects include dispersion (quantified by dispersion measure, DM), scattering (quantified by scattering time, $\tau_{\rm scat}$), scintillation (quantified by scintillation bandwidth, $\nu_{\rm scint}$), and Faraday rotation (quantified by rotation measure, RM). Repeating FRBs sometimes show a time-frequency drift in their emission patterns --- commonly referred to as the `sad trombone' effect --- which is likely a radius-to-frequency mapping phenomenon as the burst propagates outwards from the central engine \citep{2019ApJ...876L..23H}. A larger sample of FRBs at different time and frequency resolutions over large spectral extents will provide further insight into both the local environments of these sources, as well as their intrinsic emission mechanisms. This will be crucial for understanding the so-far-observed differences in time and frequency structure seen between repeating and thus-far non-repeating sources.

\textbf{Prompt multi-wavelength counterparts}: Contemporaneous optical, X-ray, or $\gamma$-ray emission accompanying an FRB would provide strong clues about the emission mechanism and the total energetics. So far, this has only been achieved with a simultaneous FRB-like radio/X-ray burst from SGR~1935+2154 whose extreme proximity compared to extragalactic FRBs made it much easier to detect \citep{brb+20, abb+20, rsf+21, msf+20, 2021NatureInsightSGR}. Multi-wavelength campaigns of even relatively nearby extragalactic FRBs have so far only resulted in upper limits \citep{scholz2020, 2023A&A...676A..17T, 2024ApJ...974..170C, 2025NatAs...9..111P, 2025Eppel, 2025MNRAS.538.1800H, Tian_2025_MNRAS}. Next generation, high-energy instruments should continue multi-wavelength observations of repeating FRBs, with particular focus on local-Universe sources. Radio instruments such as the SKA can also trigger on external multi-wavelength alerts such as those from gravitational waves, gamma-ray bursts, and neutrinos. For more details on this, see the chapter on "Rapid response triggering for radio transients with the SKA" \citep{GemmaAnderson01.2026.SKA}.

\textbf{Host galaxies}: Identification of an FRB's host galaxy requires roughly arcsecond-level localisation precision, and thus is only achievable using a radio interferometer \citep{2017ApJ...849..162E}. Once a host galaxy is known, its redshift provides a distance measurement and hence better constrains FRB energetics. However, obtaining arcsecond-level localizations for FRBs has been a major challenge, with the majority of sources detected so far localized to no better than arcminute precision \citep{chimefrbcatalog1, SecondCHIMEFRBCatalog}. Nonetheless, if an FRB's host galaxy can be determined, the global properties of the galaxy --- in terms of star-formation history, total stellar mass, etc.\ --- can also provide hints as to the nature of the FRB, similar to how the hosts of short and long GRBs have been studied \citep{2020ApJ...895L..37B,2020ApJ...903..152H, mfs+21, 2023ApJ...954...80G}. A better understanding of the host contribution enables a potential statistical reconstruction of the electron density profile in hosts and the distribution of FRBs within them \citep{2022ApJ...931...87O, 2025ApJ...991L..25L}. This can be very valuable when studying the cosmic distribution of baryons and the effect of feedback mechanisms in a cosmological setting. High-sensitivity observations that discover FRBs that are either under-luminous or very distant may also reveal new trends in the population. 

\begin{figure}
    \centering
    \includegraphics[width=1\linewidth]{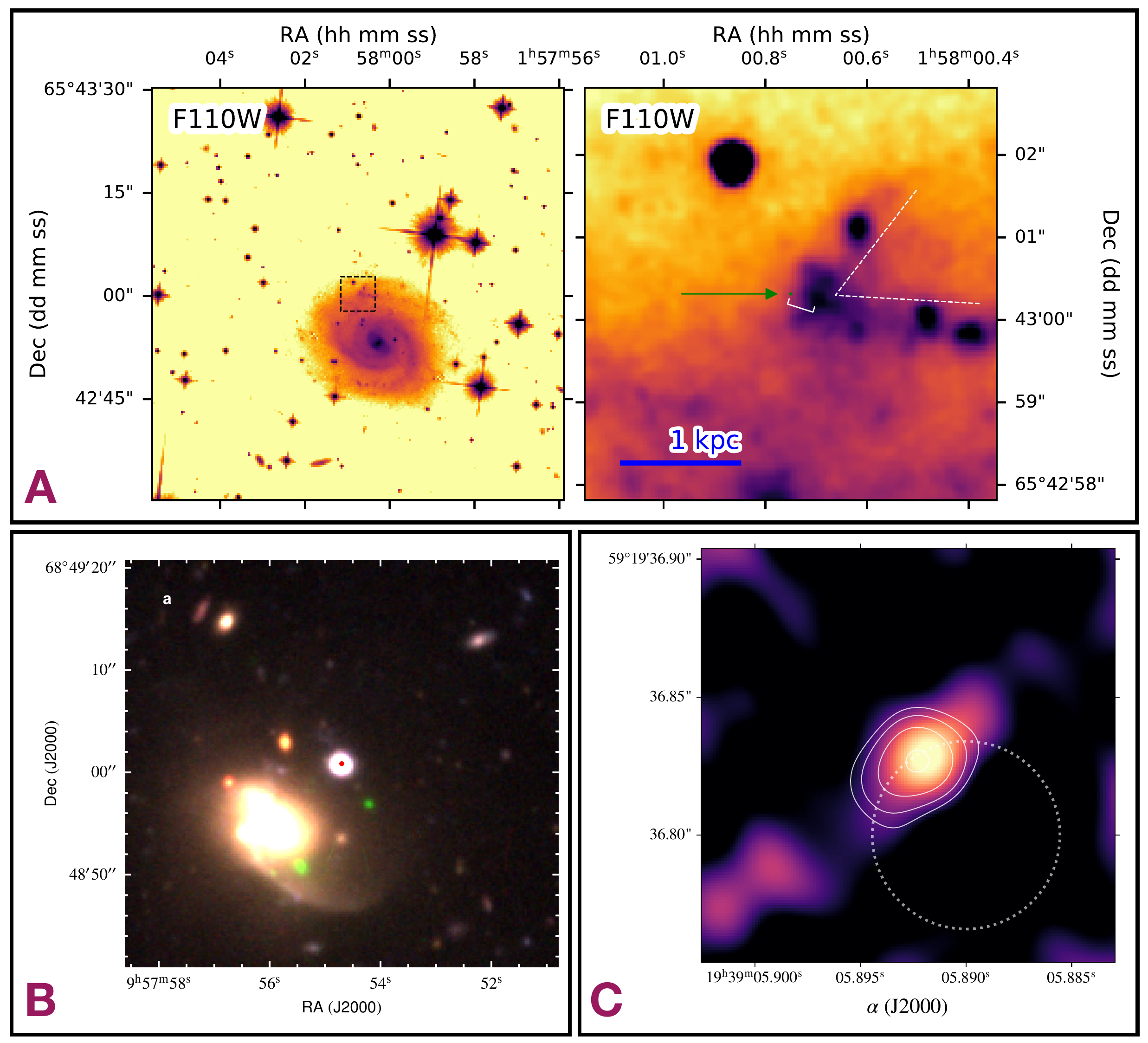}
    \caption{Three repeating FRB sources and their local environments. Panel~{\bf A}: Mas-level localisation for FRB 20180916B using the EVN overlaid on HST imaging of the field \citep{2020Natur.577..190M, 2021ApJ...908L..12T}. A zoom-in of the FRB localization within the spiral galaxy is shown in the right panel, with the FRB localization indicated with a green ellipse. Using the HST imaging, the FRB is found to be offset by $\sim200$\,pc from the peak of a nearby knot of star formation (separation indicated using a white bracket). Panel~{\bf B}: Localization of FRB~20200120E using the EVN in combination with Subaru imaging pinpoints the FRB to a globular cluster in the M81 galactic system \citep{2022Natur.602..585K}.While originally discovered using CHIME/FRB \citep{Bhardwaj_2021}, it was only through the mas-level localization that the globular cluster origin definitively determined. Panel~{\bf C}: A mas-level localisation of FRB~20190417A using the EVN (colour scale) and an associated compact persistent radio source \citep[contours;][]{moroianu2025}. EVN observations were required to confirm the compact nature of the persistent radio source ($<23$ parsec), as well as its co-location with the FRB ($<26$ parsec). For further details on these figures, see the original papers.}
    \label{fig:localisations}
\end{figure}

\textbf{Local environments}: When an FRB can be localised to milliarcsecond precision using long-baseline interferometry, then the (tens of) parsec-level local environment can be identified \citep[Figure \ref{fig:localisations};][]{2021ApJ...908L..12T, 2022Natur.602..585K, RBFloat}. This local environment can be studied in terms of star-formation rate, metallicity, and counterparts. A small number of active repeating FRBs (between three and five claimed) are associated with compact (milliarcsecond/parsec-scale) persistent radio sources, which may represent a nebula powered by the burst engine \citep[see Panel C of Figure \ref{fig:localisations};][]{2017ApJ...834L...8M, 2022Natur.606..873N, moroianu2025}. Time variation of the FRB propagation effects discussed above also reveals the conditions in the local environment by constraining its density, distribution, and magnetisation. With less than ten FRBs currently localised to a few milliarcsecond precision (equivalent to roughly tens of parsec scales in the host galaxies), a larger sample is needed to identify emerging trends and groupings.

\textbf{Redshift Evolution}:
Amongst the proposed FRB progenitor models, there is a wide variance in their expected redshift distributions. Even for models proposing a magnetar progenitor, the varying delay times between core-collapse and compact merger formation channels lead to drastically different redshift distributions for the emitted FRBs \citep{2021MNRAS.501..157Z, 2022MNRAS.510L..18J, 2024ApJ...967...29L}. As a result, understanding the redshift distribution of FRB sources can directly inform our understanding of their progenitors and their formation histories. To date, a vast majority of FRBs with known redshifts occur below a redshift of one where the impact of any source evolution in redshift is marginal. A sample of higher-redshift FRBs would be more sensitive to differences in source evolution and provide a direct way to distinguish between potential progenitors. They will also be crucial for work using FRBs as cosmological probes \citep[see accompanying chapters; ][]{Caleb01.2026.SKA, Caleb02.2026.SKA}.

\begin{figure}
    \centering
    \includegraphics[width=0.9\linewidth]{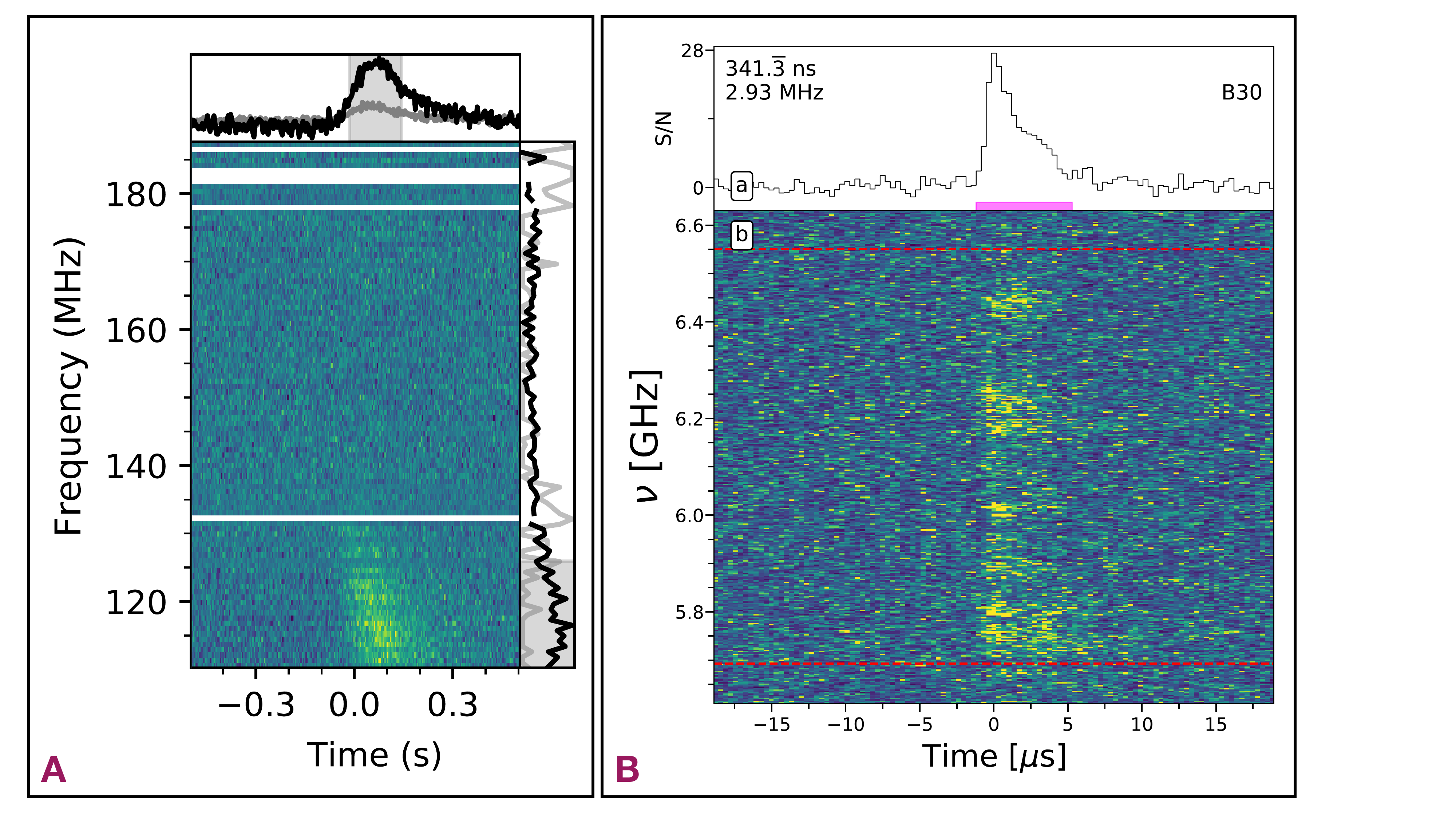}
    \caption{Panel~{\bf A}: LOFAR detection of FRB~20180916B, demonstrating emission down to the lowest-ever-detected radio frequency of 110 MHz \citep{2021ApJ...911L...3P}. Panel~{\bf B}: GBT detection of FRB~20121102A, demonstrating an isolated burst lasting only about 5 microseconds \citep{2023NatAs...7.1486S}. For further details on these figures, see the original papers. }
    \label{fig:frequencies}
\end{figure}

\section{The role of the SKA in understanding FRBs and their astrophysics}

Many telescopes worldwide have ongoing FRB observation programs. Wide-field FRB discovery systems include ASKAP/CRAFT \citep{Shannon_2025}, CHIME/FRB \citep{2018ApJ...863...48C}, DSA-110 \citep{2024ApJ...967...29L}, and MeerKAT/MeerTRAP \citep{2022MNRAS.514.1961R}. Together, these are discovering several FRBs per day, many of which can now be localised to (sub)-arcsecond precision and associated with a host galaxy. Other radio telescopes --- from small single dishes to large interferometers --- provide valuable follow-up of repeating sources across a broad frequency range with high sensitivity, cadence, and/or localisation precision. These facilities include Effelsberg, EVN, FAST, GBT, uGMRT, LOFAR, LWA, \textit{Murriyang} (Parkes), MWA, Nan\c{c}ay, NenuFAR, Onsala, Stockert, Toru\'n, VLA, VLBA, and Westerbork. There are also major FRB facilities on the horizon: in the $300-2000$\,MHz range, BURSTT \citep{2022PASP..134i4106L}, CHORD \citep{2019clrp.2020...28V}, and DSA \citep{2019BAAS...51g.255H} will expand the FRB discovery rate to at least dozens of sources per day while also providing (sub-)arcsecond-level positions.

We expect that the SKA's high instantaneous sensitivity, broad spectral coverage, large `grasp' (product of field-of-view and sensitivity) together with its frequency agility and sub-arraying capabilities will give it unique and complementary capabilities for FRB science. We outline these below.

\subsection{SKA-Low}

\textbf{Discovery Rate:}
SKA-Low could detect up to thousands of FRBs per year, comparable to the detection rates predicted for future Northern Hemisphere surveys such as CHORD and the DSA \citep[see the accompanying chapters on cosmology for further details; ][]{Caleb01.2026.SKA}. Compared to LOFAR, which is currently the world's most sensitive wide-field telescope operating in the SKA-Low band, SKA-Low~AA* will provide over an order-of-magnitude increase in sensitivity and grasp --- assuming that commensal FRB searches run with high-cadence. SKA-Low~AA4's increase in sensitivity could have a large effect on FRB detection rates as they scale non-linearly with fluence; some repeaters have burst fluence ($\mathcal{F}$ distributions as steep as $\mathcal{F}^{-2} - \mathcal{F}^{-3}$), meaning that even a $\times2$ increase in sensitivity can lead to a $\times4$ or $\times8$ increase in detection rate. However, as discussed below, factors such as scattering may play a major role in non-detections at these lower frequencies.

\textbf{Population of Low-Frequency FRBs:} FRB emission below the CHIME band ($<400$\,MHz) has been challenging to detect and hence the FRB rate in SKA-Low's spectral range is poorly understood. Despite previous searches with the GBT, uGMRT, LOFAR, LWA, MWA, and NenuFAR, only two FRB sources have been detected below $300$\,MHz \citep[][Gopinath et al. \textit{submitted}]{2021Natur.596..505P, 2021ApJ...911L...3P}. Additionally, these two sources are both repeaters and hence were only detected thanks to intense monitoring with LOFAR. Scattering is likely a major hindrance here -- an FRB that is scattered by $1$\,ms at $600$\,MHz would be broadened by about $256$\,ms at $150$\,MHz (assuming a $\nu^{-4}$ scaling), washing it out. Furthermore, the dense plasma environments of (some) FRBs may also hinder detection at low frequencies because of free-free absorption, scaling as $\nu^{-2.1}$.

Nonetheless, FRBs are coherent radio emitters and many coherent emission mechanisms produce increasing brightness towards low radio frequencies. For example, some radio pulsars have power-law spectra scaling as steeply as $\nu^{-2} - \nu^{-3}$ and the electron cyclotron maser instability (ECMI) mechanism often operates at low radio frequencies. SKA-Low's high sensitivity and large grasp from $50-350$\,MHz could thus enable a world-leading search and characterisation of FRBs at the bottom of the radio band through both high-cadence commensal observations as well as targeted monitoring of known FRB sources. SKA-Low also has the potential to be the first instrument to discover FRBs at low frequencies through untargeted searches. 

SKA-Low will nicely bridge the gap between LOFAR ($10-240$ MHz) where scattering may dominate and the CHIME band ($400-800$ MHz), where emission is often seen to the bottom of the 400 MHz band \citep{chimefrbcatalog1, SecondCHIMEFRBCatalog}. Even if FRB detectability is suppressed towards low radio frequencies, it will map how the FRB event rate decreases due to scattering and free-free absorption, giving valuable information about FRB plasma environments. 

\textbf{Simultaneous Observations with SKA-Mid:}
For sources within the joint field-of-view of SKA-Low and SKA-Mid (e.g., circumpolar FRBs), SKA-Low will be able to both shadow SKA-Mid observations through joint SKA telescope projects and trigger on SKA-Mid detections. Due to the long dispersion delay at lower frequencies, an FRB with a DM of $500$ pc cm$^{-3}$ \citep[the median DM from a catalogue of 4500 FRBs; ][]{2026arXiv260109399F} detected at 1.5 GHz will be detected at 300 MHz approximately 20-s later. Given SKA-Low's large transient buffer (up to $\sim$900-s), it should be able to easily trigger on these bursts, providing an extra 300 MHz of bandwidth coverage at lower frequencies. This will provide more information on propagation effects such as scattering and Faraday rotation measures, a further understanding of the source's emission bandwidth, and possibly an improved localization.

\textbf{Host Galaxies \& High-Redshift Sources:} Given SKA-Low's maximum baseline of 65 km, we can expect the localization capability to approach $\sim$250 mas for a 10$\sigma$ detection at 200 MHz if data from the full-array is stored within the transient buffer and calibration solutions can be applied with sufficient accuracy to limit systematic astrometric errors to below this level. With such a localization, we will be able to associate most FRBs with their host galaxies and obtain at least photometric redshifts for these sources. This is a significant benefit as the majority of discovered FRBs do not have associated host galaxies and hence redshifts \citep[although this is changing with development of instruments such as the CHIME/FRB Outriggers; ][]{2025ApJ...993...55C}. 

SKA is unique compared to CHORD and DSA in its large overlapping sky coverage with the Rubin Observatory \citep[][]{2019ApJ...873..111I}. Indeed, the accompanying chapter on FRBs as cosmological probes predicts that $>90\%$ of SKA-Low hosts will be visible with the Rubin Observatory Legacy Survey of Space and Time \citep[LSST; ][]{Caleb02.2026.SKA} . 

FRBs successfully detected by SKA-Low will also likely reside in relatively clean plasma environments. SKA-Low can thus provide an excellent sample of FRBs with host galaxies and redshifts to be used as probes. Furthermore, given we know FRBs emit in the GHz range, SKA-Low will be a key instrument for discovering high-redshift sources, as long as they are not scattered to undetectable levels by intervening galactic halos along the line-of-sight \citep{2022ApJ...934...71O}.

Localizations of order 250 mas should also give insight into where FRBs reside \textit{within} a given host galaxy. These offsets, as well as the broader host galaxy properties, can be compared with those of known and better-understood transients such as core-collapse supernovae, type Ia supernovae, short gamma-ray bursts, and long gamma-ray bursts to gain insight into the possible progenitors, as well as their formation channels \citep{2020ApJ...903..152H}. Additionally, with at least one FRB localized to a globular cluster \citep{Bhardwaj_2021, 2022Natur.602..585K}, FRB offsets can give insight into the fraction that may be associated with older stellar populations or delayed formation channels  \citep{2025ApJ...993..119G}.
Different formation channels should also have distinct redshift distributions. For example, magnetars produced through core-collapse supernovae should follow the star formation redshift rate and peak near redshifts of $z=2$, while merger-driven formation channels would have flatter redshift distributions for $0<z<2$ \citep{2022ApJ...940L..18Z}.

\textbf{Local Environment \& Persistent Radio Sources}: SKA-Low can provide valuable insight into the local environment of FRBs, and in particular the persistent radio emission that is associated with a rare few but highly interesting, FRBs. Higher-frequency work has shown that the spectral index of these sources is flat in the range of 745 to 1400 MHz \citep{2023MNRAS.525.3626R, 2025arXiv250623861B}. However, current instruments lack the sensitivity to probe these sources at lower frequencies \citep[limits in the range of $\sim200\mu$Jy at 400 MHz with uGMRT for FRB 20121102A; ][]{2025arXiv250623861B}. With imaging sensitivity down to the $\mu$Jy-level, SKA-Low can test whether there is a spectral turnover in the persistent radio emission and hence provide insight into the astrophysical mechanisms driving them. Detections at these frequencies will also have direct implications for the density of the local environment. However, separation of the compact emission associated with the FRB from other low-frequency radio emission associated with the host galaxy will generally require higher angular resolution than can be provided by SKA-Low alone, necessitating the use of SKA-Low VLBI. For more  details on SKA-Low VLBI, see the chapter on "Low-frequency VLBI with the SKA-Low" \citep{Timmerman01.2026.SKA}.

\subsection{SKA-Mid}

\textbf{Discovery Rate:} SKA-Mid is predicted to detect tens to hundreds of FRBs per year, on par with current FRB experiments such as CHIME/FRB \citep{Caleb02.2026.SKA}. Importantly, SKA-Mid's excellent sensitivity will give it access to many high-redshift ($z > 1$) FRBs per year. The SKA-Mid precursor MeerKAT has already given us a glimpse in this direction through its real-time commensal FRB system, MeerTRAP \citep{2022MNRAS.514.1961R}. In particular, \citet{2025arXiv250801648C} presented the record-breaking discovery of an FRB at $z = 2.15$ using MeerKAT, doubling the redshift of the most distant FRB yet discovered. 

\textbf{FRB Population at GHz Frequencies:} The vast majority of FRBs, now thousands of sources, have been discovered in the $400-800$\,MHz range thanks to the unparalleled $200 \: \text{deg}^2$ field-of-view and powerful real-time processing infrastructure of the CHIME/FRB system \citep{chimefrbcatalog1, SecondCHIMEFRBCatalog}. A sizeable population of hundreds of FRBs have also been discovered in the $800-2400$\,MHz range using ASKAP \citep{Shannon_2025}, DSA-110 \citep{2024ApJ...967...29L}, and MeerKAT \citep{2023MNRAS.524.4275J, 2025arXiv250705982P}. Thus, we have a reasonably good picture of the observational properties and event rates of FRBs in the spectral ranges covered by SKA-Mid Band~1 ($350-1050$\,MHz) and Band~2 ($950-1760$\,MHz). In contrast, little is known about the FRB emission and event rates in the higher-frequency ranges covered by SKA-Mid Bands~5a/5b ($4600-8500$\,MHz/$8300-15400$\,MHz) --- though targeted follow-up of repeating sources has confirmed that FRBs emit in this range \citep{2018ApJ...863....2G, 2025arXiv251008367L} and at least one thus-far one-off FRB has been detected at S-band with MeerKAT \citep{2025arXiv250705982P}.

SKA-Mid Bands~5a/5b have the unique potential to greatly expand our understanding of the FRB emission at high radio frequencies. FRBs are harder to discover at such frequencies because of the reduced field-of-view and potentially negative spectral index. However, propagation effects like scattering and dispersion have a negligible effect on detectability at frequencies above $5$\,GHz, so there is potentially a population of FRBs in high-density environments that can only be discovered using SKA-Mid Bands~5a/5b. Here too, commensal searches leveraging as many SKA-Mid observations as possible will be crucial for success. 

By searching for bursts at high time resolutions and by saving voltage/complex channelised data in its transient buffer, SKA-Mid can also chart a largely unexplored parameter space of ultra-fast radio bursts with timescales of only microseconds \citep{2023NatAs...7.1486S, 2022NatAs...6..393N}. Indeed, SKA-Mid's FRB search time resolution at $\sim$64$ \mu$s is approximately $\times5$ higher than that of MeerKAT and will be one of the highest time resolution searches for FRBs. SKA-Mid may also detect an intrinsically "not-so-fast" population of FRBs with little scattering yet wide burst widths that would be difficult to detect at lower frequencies due to the reduced peak flux from scattering.

SKA-Mid may discover a sub-class of FRBs at higher frequencies with different progenitors. While compact object merger rates cannot explain all FRBs (and their cataclysmic nature is inconsistent with repeating sources), \citet{MostPhilippov2020, 2022MostPhilippov, 2023MostPhilippov} predict a population of FRBs in the 10 to 20 GHz range from interactions in the current sheets formed prior to a compact object merger. Given the expected dense ejecta from a merger or a supernova, high-frequency SKA-Mid follow-up will also offer the best opportunity for detecting pulses from a millisecond magnetar formed post merger/explosion. Such detections would provide the strongest evidence yet for FRBs formed through one of these channels. Additionally, SKA-Mid may bridge the gap between the FRB-like bursts from SGR 1935+2154 and extragalactic FRBs through targeted searches for low-luminosity, local Universe FRBs \citep{2023A&A...674A.223P, 2024MNRAS.528.6340P}. These local Universe sources will likely be the best candidates for multi-wavelength counterpart searches. 

\textbf{High-Redshift FRBs:} With increased sensitivity and GHz frequency ranges, SKA-Mid can discover many high-redshift (z$>1$) FRBs that are either too faint or too scattered to be detected at lower frequencies and sensitivities.  A larger sample of high-redshift sources will help us track the volumetric rate of FRBs as a function of cosmic time, giving clues to their progenitors. High-redshift FRBs are also typically the most energetic. Thus, we can use them to place strong constraints on the total energy budget of FRBs, informing both the emission mechanisms and possible progenitors. It is also possible that more extreme FRB-like emitters could be discovered in the early Universe, and that they would have different progenitors than those at lower redshifts. SKA-Mid could thus expand our view of the first FRB sources to form in the Universe.

\textbf{Repeating \& Non-repeating FRBs:} Current measurements for non-repeater bandwidths are often limited by the spectral extent of the detection instrument, with most non-repeating CHIME/FRB bursts spanning the full 400-MHz bandwidth and hence only providing a lower limit on the total emission bandwidth \citep{2021ApJ...911L...3P, scm+24}. Pulsar pulses, on the other hand, can have spectral extents that can span over 8 GHz \citep{2012A&A...543A..66H}. While the maximum transient buffer bandwidth for SKA-Mid is currently 400 MHz, simultaneous observations with SKA-Low could test the full emission bandwidth of the thus-far non-repeating FRB sources. Indeed, a significant fraction of the sky ($20-30\%$) should be simultaneously visible to both SKA-Low and SKA-Mid. An understanding of the intrinsic non-repeater bandwidths will have direct implications for the assumed energy budget of non-repeating sources, and may further exacerbate the difference seen between the spectral structure of repeating (narrrow-band) and non-repeating (broadband) FRBs. 

Beyond discovering new sources, SKA-Mid will also be an excellent tool for follow-up observations. With increased sensitivity, the SKA
will be able search for ultra-faint repeat bursts at non-repeater locations, testing the hypothesis that all FRBs eventually repeat. 
As the most sensitive planned radio telescope in the Southern Hemisphere, SKA will be unique in its ability to perform this follow-up. Additionally, using the pulsar timing beams with 2.5 GHz of frequency coverage, SKA-Mid can search for temporally simultaneous repeat bursts at different portions of the FRB spectrum. While searches for this at known repeater locations has been performed \citep{2021MNRAS.500.2525K, 2023MNRAS.524.3303B, 2025arXiv251008367L}, it has been limited to frequencies $<10$ GHz, has only been done for a small number of repeaters, and has not always been continuous in frequency coverage. Observations using the sub-arraying capabilities of MeerKAT have already demonstrated the potential of SKA-Mid in this area -- with a continuous frequency coverage from 544 to 1756\,MHz, it found complex burst frequency shapes and evolution from a repeating FRB \citep{Tian_2025_MNRAS}.

\textbf{Host Galaxies:} Thanks to the combination of a sensitive dense core and long baselines in a spiral layout, SKA-Mid will be both an FRB detection and localisation machine. With AA* we can expect a nominal localisation accuracy of $\sim75\,\mathrm{mas}$
for a $10\sigma$ detection at an observing frequency of $1.4\,\mathrm{GHz}$, assuming diffraction-limited astrometric accuracy. Such a localization is more than sufficient to associate the FRB with a host galaxy and, in the case of nearby FRBs, can even enable the characterisation of the local environment with localizations of order tens to hundreds of parsecs. Once AA4 becomes available, the localisation precision of SKA-Mid alone will be of order $15\,\mathrm{mas}$. SKA is unique in its significant field-of-view overlap with Rubin Observatory as compared to Northern Hemisphere radio instruments. The accompanying chapter on FRBs as cosmological probes predicts that for both Band 1 and Band 2 for SKA-Mid in AA*, approximately $80\%$ of FRB hosts will be visible with LSST \citep{Caleb01.2026.SKA}. Thus, for AA4, we can routinely assign host galaxies \textit{and} study the local environment of nearby and moderately distant FRBs.

\textbf{Local Environments \& Persistent Radio Sources:} Coupling SKA-Mid to VLBI networks like the EVN or the LBA will provide milliarcsecond localisations of repeating sources that can be compared with optical imaging from ELT, HST, and JWST (see, e.g., Figure \ref{fig:localisations}). This will allow us to zoom in on FRB progenitor environments on scales comparable to the distance between stars in the host galaxy; we may even be able to associate an FRB with a specific companion star, if it is sufficiently massive and luminous. Indeed, with a 13-pc localization for FRB 20250316A \citep{RBFloat}, JWST observations were able to directly search for a magnetar at the FRB location  \citep{2025ApJ...989L..49B}. 

SKA-Mid VLBI will also probe persistent radio emission at the locations of FRBs. This emission likely serves as a calorimeter of the central engine's cumulative energy output, and can thus inform the possible FRB progenitors \citep{2017ApJ...841...14M, Margalit_2018}. SKA-Mid VLBI can constrain the compactness of the persistent radio source and characterise its broad-band spectrum. For more  details on SKA-Mid VLBI, see the chapter on "Extending the SKA Across Africa: The Case for a Continental African VLBI Network" \citep{Bempong-Manful01.2026.SKA}.

\section{Looking forward}
By the early 2030s, CHORD, the DSA, and the SKA should be finding tens of thousands of new FRB sources per year. To remain competitive in this new era, the SKA must prioritize significant commensal searches that enable high numbers of FRB detections in new parts of the transient parameter space. With its large frequency extent from 50 MHz (SKA-Low) up to 15 GHz (SKA-Mid) and its high time resolution searches (e.g., search timescales of $\sim64\mu$s), it has the potential to greatly extend the transient phase space along the frequency and time axes. It will also nicely complement future experiments such as DSA and CHORD, offering similar sensitivity but in the Southern Hemisphere. And, with significant overlapping sky coverage with the Rubin Observatory, it will be able to provide a large sample of FRB host galaxies and hence redshifts which will be beneficial both for understanding their progenitors and formation channels, as well as for cosmology.

\newpage
\section*{Acknowledgments}
We dedicate this chapter to the memory of J.P. Macquart, who wrote the original SKA FRB Science Book chapter ``Fast Transients at Cosmological Distances with the SKA''.

We thank Amanda Cook, Alex Cooper, Adam Deller, Maura Pilia, and Lauren Rhodes for their helpful comments and insights.

A.P.C. is a Canadian SKA Scientist and is funded by the Government of Canada / est financé par le gouvernement du Canada. The AstroFlash research group at McGill University, University of Amsterdam, ASTRON, and JIVE is supported by: a Canada Excellence Research Chair in Transient Astrophysics (CERC-2022-00009); an Advanced Grant from the European Research Council (ERC) under the European Union’s Horizon 2020 research and innovation programme (`EuroFlash'; Grant agreement No. 101098079); and an NWO-Vici grant (`AstroFlash'; VI.C.192.045). M.W.S is a Fonds de Recherche du Qu\'ebec - Nature et Technologies (FRQNT) postdoctoral fellow and acknowledges support from the Trottier Space Institute Fellowship program. L.G.S is a Lise Meitner Research Group leader and acknowledges funding
from the Max Planck Society.

\bibliographystyle{abbrvnat-maxbibnames4}
\bibliography{chapter} %

\begin{thebibliography}{115}
\providecommand{\natexlab}[1]{#1}
\providecommand{\url}[1]{\texttt{#1}}
\expandafter\ifx\csname urlstyle\endcsname\relax
  \providecommand{\doi}[1]{doi: #1}\else
  \providecommand{\doi}{doi: \begingroup \urlstyle{rm}\Url}\fi

\bibitem[Anderson et~al.(2026)Anderson, author2, author3, author4, and author5]{GemmaAnderson01.2026.SKA}
G.~E. Anderson et al.
\newblock In \emph{Advancing Astrophysics with the SKA -- II (AASKAII)}. 2026.
\newblock arXiv search: Report number AASKAII/GemmaAnderson01.

\bibitem[{Anna-Thomas} et~al.(2023){Anna-Thomas}, {Connor}, {Dai}, {Feng}, {Burke-Spolaor}, {Beniamini}, {Yang}, {Zhang}, {Aggarwal}, {Law}, {Li}, {Niu}, {Chatterjee}, {Cruces}, {Duan}, {Filipovic}, {Hobbs}, {Lynch}, {Miao}, {Niu}, {Ocker}, {Tsai}, {Wang}, {Xue}, {Yao}, {Yu}, {Zhang}, {Zhang}, {Zhu}, and {Zhu}]{2023Sci...380..599A}
R.~{Anna-Thomas} et al.
\newblock \emph{Science}, 380\penalty0 (6645):\penalty0 599--603, May 2023.
\newblock \doi{10.1126/science.abo6526}.

\bibitem[{Beloborodov}(2020)]{2020ApJ...896..142B}
A.~M. {Beloborodov}.
\newblock \emph{\apj}, 896\penalty0 (2):\penalty0 142, June 2020.
\newblock \doi{10.3847/1538-4357/ab83eb}.

\bibitem[Bempong-Manful et~al.(2026)Bempong-Manful, author2, author3, author4, and author5]{Bempong-Manful01.2026.SKA}
E.~Bempong-Manful et al.
\newblock In \emph{Advancing Astrophysics with the SKA -- II (AASKAII)}. 2026.
\newblock arXiv search: Report number AASKAII/Bempong-Manful01.

\bibitem[{Berger}(2014)]{2014Berger}
E.~{Berger}.
\newblock \emph{\araa}, 52:\penalty0 43--105, Aug. 2014.
\newblock \doi{10.1146/annurev-astro-081913-035926}.

\bibitem[{Bethapudi} et~al.(2023){Bethapudi}, {Spitler}, {Main}, {Li}, and {Wharton}]{2023MNRAS.524.3303B}
S.~{Bethapudi} et al.
\newblock \emph{\mnras}, 524\penalty0 (3):\penalty0 3303--3313, Sept. 2023.
\newblock \doi{10.1093/mnras/stad2009}.

\bibitem[{Bhandari} et~al.(2020){Bhandari}, {Sadler}, {Prochaska}, {Simha}, {Ryder}, {Marnoch}, {Bannister}, {Macquart}, {Flynn}, {Shannon}, {Tejos}, {Corro-Guerra}, {Day}, {Deller}, {Ekers}, {Lopez}, {Mahony}, {Nu{\~n}ez}, and {Phillips}]{2020ApJ...895L..37B}
S.~{Bhandari} et al.
\newblock \emph{\apjl}, 895\penalty0 (2):\penalty0 L37, June 2020.
\newblock \doi{10.3847/2041-8213/ab672e}.

\bibitem[Bhardwaj et~al.(2021)Bhardwaj, Gaensler, Kaspi, Landecker, Mckinven, Michilli, Pleunis, Tendulkar, Andersen, Boyle, Cassanelli, Chawla, Cook, Dobbs, Fonseca, Kaczmarek, Leung, Masui, Mnchmeyer, Ng, Rafiei-Ravandi, Scholz, Shin, Smith, Stairs, and Zwaniga]{Bhardwaj_2021}
M.~Bhardwaj et al.
\newblock \emph{The Astrophysical Journal Letters}, 910\penalty0 (2):\penalty0 L18, Mar. 2021.
\newblock ISSN 2041-8213.
\newblock \doi{10.3847/2041-8213/abeaa6}.
\newblock URL \url{http://dx.doi.org/10.3847/2041-8213/abeaa6}.

\bibitem[{Bhardwaj} et~al.(2021){Bhardwaj}, {Kirichenko}, {Michilli}, {Mayya}, {Kaspi}, {Gaensler}, {Rahman}, {Tendulkar}, {Fonseca}, {Josephy}, {Leung}, {Merryfield}, {Petroff}, {Pleunis}, {Sanghavi}, {Scholz}, {Shin}, {Smith}, and {Stairs}]{bkm+21}
M.~{Bhardwaj} et al.
\newblock \emph{ApJL}, 919\penalty0 (2):\penalty0 L24, Oct. 2021.
\newblock \doi{10.3847/2041-8213/ac223b}.

\bibitem[{Bhardwaj} et~al.(2025){Bhardwaj}, {Balasubramanian}, {Kaushal}, and {Tendulkar}]{2025arXiv250623861B}
M.~{Bhardwaj}, A.~{Balasubramanian}, Y.~{Kaushal}, and S.~P. {Tendulkar}.
\newblock \emph{arXiv e-prints}, art. arXiv:2506.23861, June 2025.
\newblock \doi{10.48550/arXiv.2506.23861}.

\bibitem[{Blanchard} et~al.(2025){Blanchard}, {Berger}, {Andrew}, {Suresh}, {Uno}, {Kilpatrick}, {Metzger}, {Kumar}, {Sridhar}, {Cook}, {Dong}, {Eftekhari}, {Fong}, {Golay}, {Hiramatsu}, {Joseph}, {Kaspi}, {Lazda}, {Leung}, {Masui}, {Mena-Parra}, {Nimmo}, {Pearlman}, {Shah}, {Shin}, and {Simha}]{2025ApJ...989L..49B}
P.~K. {Blanchard} et al.
\newblock \emph{\apjl}, 989\penalty0 (2):\penalty0 L49, Aug. 2025.
\newblock \doi{10.3847/2041-8213/adf29f}.

\bibitem[{Bochenek} et~al.(2020){Bochenek}, {Ravi}, {Belov}, {Hallinan}, {Kocz}, {Kulkarni}, and {McKenna}]{brb+20}
C.~D. {Bochenek} et al.
\newblock \emph{\nat}, 587\penalty0 (7832):\penalty0 59--62, Nov. 2020.
\newblock \doi{10.1038/s41586-020-2872-x}.

\bibitem[{Brandenberger} et~al.(2017){Brandenberger}, {Cyr}, and {Varna Iyer}]{2017arXiv170702397B}
R.~{Brandenberger}, B.~{Cyr}, and A.~{Varna Iyer}.
\newblock \emph{arXiv e-prints}, art. arXiv:1707.02397, July 2017.
\newblock \doi{10.48550/arXiv.1707.02397}.

\bibitem[{Caleb} et~al.(2025){Caleb}, {Nanayakkara}, {Stappers}, {Pastor-Marazuela}, {Khrykin}, {Glazebrook}, {Tejos}, {Prochaska}, {Rajwade}, {Mas-Ribas}, {Driessen}, {Fong}, {Gordon}, {Hoffmann}, {James}, {Jankowski}, {Kahinga}, {Kramer}, {Simha}, {Barr}, {Christiaan Bezuidenhout}, {Deng}, {Lin}, {Marnoch}, {Martin}, {Nugent}, {Shaji}, and {Tian}]{2025arXiv250801648C}
M.~{Caleb} et al.
\newblock \emph{arXiv e-prints}, art. arXiv:2508.01648, Aug. 2025.
\newblock \doi{10.48550/arXiv.2508.01648}.

\bibitem[Caleb et~al.(2026{\natexlab{a}})Caleb, author2, author3, author4, and author5]{Caleb01.2026.SKA}
M.~Caleb et al.
\newblock In \emph{Advancing Astrophysics with the SKA -- II (AASKAII)}. 2026{\natexlab{a}}.
\newblock arXiv search: Report number AASKAII/Caleb01.

\bibitem[Caleb et~al.(2026{\natexlab{b}})Caleb, author2, author3, author4, and author5]{Caleb02.2026.SKA}
M.~Caleb et al.
\newblock In \emph{Advancing Astrophysics with the SKA -- II (AASKAII)}. 2026{\natexlab{b}}.
\newblock arXiv search: Report number AASKAII/Caleb02.

\bibitem[{Chatterjee} et~al.(2017){Chatterjee}, {Law}, {Wharton}, {Burke-Spolaor}, {Hessels}, {Bower}, {Cordes}, {Tendulkar}, {Bassa}, {Demorest}, {Butler}, {Seymour}, {Scholz}, {Abruzzo}, {Bogdanov}, {Kaspi}, {Keimpema}, {Lazio}, {Marcote}, {McLaughlin}, {Paragi}, {Ransom}, {Rupen}, {Spitler}, and {van Langevelde}]{Chatterjee2017}
S.~{Chatterjee} et al.
\newblock \emph{\nat}, 541\penalty0 (7635):\penalty0 58--61, Jan. 2017.
\newblock \doi{10.1038/nature20797}.

\bibitem[{CHIME/FRB Collaboration}(2025)]{SecondCHIMEFRBCatalog}
{CHIME/FRB Collaboration}.
\newblock The second chime/frb catalog of fast radio bursts.
\newblock Submitted to ApJ, 2025.

\bibitem[{CHIME/FRB Collaboration} et~al.(2018){CHIME/FRB Collaboration}, {Amiri}, {Bandura}, {Berger}, {Bhardwaj}, {Boyce}, {Boyle}, {Brar}, {Burhanpurkar}, {Chawla}, {Chowdhury}, {Cliche}, {Cranmer}, {Cubranic}, {Deng}, {Denman}, {Dobbs}, {Fandino}, {Fonseca}, {Gaensler}, {Giri}, {Gilbert}, {Good}, {Guliani}, {Halpern}, {Hinshaw}, {H{\"o}fer}, {Josephy}, {Kaspi}, {Landecker}, {Lang}, {Liao}, {Masui}, {Mena-Parra}, {Naidu}, {Newburgh}, {Ng}, {Patel}, {Pen}, {Pinsonneault-Marotte}, {Pleunis}, {Rafiei Ravandi}, {Ransom}, {Renard}, {Scholz}, {Sigurdson}, {Siegel}, {Smith}, {Stairs}, {Tendulkar}, {Vanderlinde}, and {Wiebe}]{2018ApJ...863...48C}
{CHIME/FRB Collaboration} et al.
\newblock \emph{ApJ}, 863:\penalty0 48, Aug. 2018.
\newblock \doi{10.3847/1538-4357/aad188}.

\bibitem[{CHIME/FRB Collaboration} et~al.(2020){CHIME/FRB Collaboration}, {Andersen}, {Bandura}, {Bhardwaj}, {Bij}, {Boyce}, {Boyle}, {Brar}, {Cassanelli}, {Chawla}, {Chen}, {Cliche}, {Cook}, {Cubranic}, {Curtin}, {Denman}, {Dobbs}, {Dong}, {Fandino}, {Fonseca}, {Gaensler}, {Giri}, {Good}, {Halpern}, {Hill}, {Hinshaw}, {H{\"o}fer}, {Josephy}, {Kania}, {Kaspi}, {Landecker}, {Leung}, {Li}, {Lin}, {Masui}, {McKinven}, {Mena-Parra}, {Merryfield}, {Meyers}, {Michilli}, {Milutinovic}, {Mirhosseini}, {M{\"u}nchmeyer}, {Naidu}, {Newburgh}, {Ng}, {Patel}, {Pen}, {Pinsonneault-Marotte}, {Pleunis}, {Quine}, {Rafiei-Ravandi}, {Rahman}, {Ransom}, {Renard}, {Sanghavi}, {Scholz}, {Shaw}, {Shin}, {Siegel}, {Singh}, {Smegal}, {Smith}, {Stairs}, {Tan}, {Tendulkar}, {Tretyakov}, {Vanderlinde}, {Wang}, {Wulf}, and {Zwaniga}]{abb+20}
{CHIME/FRB Collaboration} et al.
\newblock \emph{\nat}, 587\penalty0 (7832):\penalty0 54--58, Nov. 2020.
\newblock \doi{10.1038/s41586-020-2863-y}.

\bibitem[{CHIME/FRB Collaboration} et~al.(2021){CHIME/FRB Collaboration}, {Amiri}, {Andersen}, {Bandura}, {Berger}, {Bhardwaj}, {Boyce}, {Boyle}, {Brar}, {Breitman}, {Cassanelli}, {Chawla}, {Chen}, {Cliche}, {Cook}, {Cubranic}, {Curtin}, {Deng}, {Dobbs}, {(Adam) Dong}, {Eadie}, {Fandino}, {Fonseca}, {Gaensler}, {Giri}, {Good}, {Halpern}, {Hill}, {Hinshaw}, {Josephy}, {Kaczmarek}, {Kader}, {Kania}, {Kaspi}, {Landecker}, {Lang}, {Leung}, {Li}, {Lin}, {Masui}, {McKinven}, {Mena-Parra}, {Merryfield}, {Meyers}, {Michilli}, {Milutinovic}, {Mirhosseini}, {M{\"u}nchmeyer}, {Naidu}, {Newburgh}, {Ng}, {Patel}, {Pen}, {Petroff}, {Pinsonneault-Marotte}, {Pleunis}, {Rafiei-Ravandi}, {Rahman}, {Ransom}, {Renard}, {Sanghavi}, {Scholz}, {Shaw}, {Shin}, {Siegel}, {Sikora}, {Singh}, {Smith}, {Stairs}, {Tan}, {Tendulkar}, {Vanderlinde}, {Wang}, {Wulf}, and {Zwaniga}]{chimefrbcatalog1}
{CHIME/FRB Collaboration} et al.
\newblock \emph{ApJs}, 257\penalty0 (2):\penalty0 59, Dec. 2021.
\newblock \doi{10.3847/1538-4365/ac33ab}.

\bibitem[{CHIME/FRB Collaboration} et~al.(2025{\natexlab{a}}){CHIME/FRB Collaboration}, {Abbott}, {Amouyal}, {Andersen}, {Andrew}, {Bandura}, {Bhardwaj}, {Bhopi}, {Bhusare}, {Brar}, {Cai}, {Cassanelli}, {Chatterjee}, {Cliche}, {Cook}, {Curtin}, {Davies-Velie}, {Dobbs}, {Dong}, {Dong}, {Eadie}, {Eftekhari}, {Fong}, {Fonseca}, {Gaensler}, {Gusinskaia}, {Hessels}, {Hewitt}, {Huang}, {Jain}, {Joseph}, {Kahinga}, {Kaspi}, {Khan}, {Kharel}, {Lanman}, {L'Argent}, {Lazda}, {Leung}, {Main}, {Mas-Ribas}, {Masui}, {McGregor}, {McKinven}, {Mena-Parra}, {Michilli}, {Mulyk}, {Ng}, {Nimmo}, {Pandhi}, {Patil}, {Pearlman}, {Pen}, {Pleunis}, {Prochaska}, {Rafiei-Ravandi}, {Ransom}, {Sachdeva}, {Sammons}, {Sand}, {Scholz}, {Shah}, {Shin}, {Siegel}, {Simha}, {Smith}, {Stairs}, {Stenning}, {Wang}, {Boles}, {Cognard}, {Dijkema}, {Filippenko}, {Gawro{\'n}ski}, {Herrmann}, {Kilpatrick}, {Kirsten}, {Knabel}, {Ould-Boukattine}, {Paugnat}, {Puchalska}, {Sheu}, {Suresh}, {Tohuvavohu}, {Treu}, and {Zheng}]{RBFloat}
{CHIME/FRB Collaboration} et al.
\newblock \emph{\apjl}, 989\penalty0 (2):\penalty0 L48, Aug. 2025{\natexlab{a}}.
\newblock \doi{10.3847/2041-8213/adf62f}.

\bibitem[{CHIME/FRB Collaboration} et~al.(2025{\natexlab{b}}){CHIME/FRB Collaboration}, {Amiri}, {Amouyal}, {Andersen}, {Andrew}, {Bandura}, {Bhardwaj}, {Boyle}, {Brar}, {Cassity}, {Chatterjee}, {Curtin}, {Dobbs}, {Dong}, {Dong}, {Eadie}, {Eftekhari}, {Fong}, {Fonseca}, {Gaensler}, {Halpern}, {Hessels}, {Hopkins}, {Ibik}, {Joseph}, {Kaczmarek}, {Kahinga}, {Kaspi}, {Khairy}, {Kilpatrick}, {Lanman}, {Lazda}, {Leung}, {Main}, {Mas-Ribas}, {Masui}, {McKinven}, {Mena-Parra}, {Meyers}, {Michilli}, {Milutinovic}, {Nimmo}, {Noble}, {Pandhi}, {Patil}, {Pearlman}, {Petroff}, {Pleunis}, {Prochaska}, {Rafiei-Ravandi}, {Rahman}, {Renard}, {Sammons}, {Sand}, {Scholz}, {Shah}, {Shin}, {Siegel}, {Simha}, {Smith}, {Stairs}, {Vanderlinde}, {Wang}, {Wulf}, and {Zegmott}]{2025ApJS..280....6C}
{CHIME/FRB Collaboration} et al.
\newblock \emph{\apjs}, 280\penalty0 (1):\penalty0 6, Sept. 2025{\natexlab{b}}.
\newblock \doi{10.3847/1538-4365/addbda}.

\bibitem[{CHIME/FRB Collaboration} et~al.(2025{\natexlab{c}}){CHIME/FRB Collaboration}, {Amiri}, {Andersen}, {Andrew}, {Bandura}, {Bhardwaj}, {Bhopi}, {Bidula}, {Boyle}, {Brar}, {Carlson}, {Cassanelli}, {Cassity}, {Chatterjee}, {Cliche}, {Curtin}, {Darlinger}, {Deboer}, {Dobbs}, {Dong}, {Eadie}, {Fonseca}, {Gaensler}, {Gusinskaia}, {Halpern}, {Hendricksen}, {Hessels}, {Joseph}, {Kaczmarek}, {Kaspi}, {Khairy}, {Landecker}, {Lanman}, {Lau}, {Lazda}, {Leung}, {Main}, {Masui}, {McKinven}, {Mena-Parra}, {Meyers}, {Michilli}, {Milutinovic}, {Nimmo}, {Noble}, {Pandhi}, {Pearlman}, {Peterson}, {Petroff}, {Pleunis}, {Pollak}, {Rafiei-Ravandi}, {Renard}, {Sammons}, {Sand}, {Sanghavi}, {Scholz}, {Shah}, {Shin}, {Siegel}, {Siemion}, {Sievers}, {Smith}, {Spear}, {Stairs}, {Vanderlinde}, {Wang}, {Willis}, and {Zegmott}]{2025ApJ...993...55C}
{CHIME/FRB Collaboration} et al.
\newblock \emph{\apj}, 993\penalty0 (1):\penalty0 55, Nov. 2025{\natexlab{c}}.
\newblock \doi{10.3847/1538-4357/adfdcc}.

\bibitem[{CHIME/FRB Collaboration} et~al.(2026){CHIME/FRB Collaboration}, {Abbott}, {Andersen}, {Andrew}, {Bandura}, {Bhardwaj}, {Bhusare}, {Brar}, {Cassanelli}, {Chatterjee}, {Cliche}, {Cook}, {Curtin}, {Dobbs}, {Dong}, {Eadie}, {Eftekhari}, {Fonseca}, {Gaensler}, {Good}, {Halpern}, {Hessels}, {Ibik}, {Jain}, {Joseph}, {Kader}, {Kaspi}, {Khan}, {Kharel}, {Kumar}, {Landecker}, {Lang}, {Lanman}, {L'Argent}, {Lazda}, {Leung}, {Li}, {Lintott}, {Main}, {Masui}, {Mate}, {McGregor}, {Mckinven}, {Mena-Parra}, {Meyers}, {Michilli}, {Ng}, {Ng}, {Nimmo}, {Noble}, {Pandhi}, {Patil}, {Pearlman}, {Pen}, {Pleunis}, {Prochaska}, {Rafiei-Ravandi}, {Ransom}, {Renard}, {Sammons}, {Sand}, {Scholz}, {Shah}, {Shin}, {Siegel}, {Sirota}, {Smith}, {Stairs}, {Stenning}, {Tendulkar}, {Vanderlinde}, {Walmsley}, {Wang}, and {Wulf}]{2026arXiv260109399F}
{CHIME/FRB Collaboration} et al.
\newblock \emph{arXiv e-prints}, art. arXiv:2601.09399, Jan. 2026.
\newblock \doi{10.48550/arXiv.2601.09399}.

\bibitem[{Cook} et~al.(2024){Cook}, {Scholz}, {Pearlman}, {Abbott}, {Cruces}, {Gaensler}, {Dong}, {Michilli}, {Eadie}, {Kaspi}, {Stairs}, {Tan}, {Bhardwaj}, {Cassanelli}, {Curtin}, {Ibik}, {Lazda}, {Masui}, {Pandhi}, {Rafiei-Ravandi}, {Sammons}, {Shin}, {Smith}, and {Stenning}]{2024ApJ...974..170C}
A.~M. {Cook} et al.
\newblock \emph{\apj}, 974\penalty0 (2):\penalty0 170, Oct. 2024.
\newblock \doi{10.3847/1538-4357/ad6a13}.

\bibitem[{Cooper} et~al.(2023){Cooper}, {Gupta}, {Wadiasingh}, {Wijers}, {Boersma}, {Andreoni}, {Rowlinson}, and {Gourdji}]{Cooper2023}
A.~J. {Cooper} et al.
\newblock \emph{MNRAS}, 519\penalty0 (3):\penalty0 3923--3946, Mar. 2023.
\newblock \doi{10.1093/mnras/stac3580}.

\bibitem[{Curtin} et~al.(2025){Curtin}, {Sand}, {Pleunis}, {Jain}, {Kaspi}, {Michilli}, {Fonseca}, {Shin}, {Nimmo}, {Brar}, {Dong}, {Eadie}, {Gaensler}, {Herrera-Martin}, {Ibik}, {Joseph}, {Kaczmarek}, {Leung}, {Main}, {Masui}, {Mckinven}, {Mena-Parra}, {Ng}, {Pandhi}, {Pearlman}, {Rafiei-Ravandi}, {Sammons}, {Scholz}, {Smith}, and {Stairs}]{2025ApJ...992..206C}
A.~P. {Curtin} et al.
\newblock \emph{\apj}, 992\penalty0 (2):\penalty0 206, Oct. 2025.
\newblock \doi{10.3847/1538-4357/adf844}.

\bibitem[{Eftekhari} and {Berger}(2017)]{2017ApJ...849..162E}
T.~{Eftekhari} and E.~{Berger}.
\newblock \emph{\apj}, 849\penalty0 (2):\penalty0 162, Nov. 2017.
\newblock \doi{10.3847/1538-4357/aa90b9}.

\bibitem[{Eftekhari} et~al.(2024){Eftekhari}, {Dong}, {Fong}, {Shah}, {Simha}, {Andersen}, {Andrew}, {Bhardwaj}, {Cassanelli}, {Chatterjee}, {Coulter}, {Fonseca}, {Gaensler}, {Gordon}, {Hessels}, {Ibik}, {Joseph}, {Kahinga}, {Kaspi}, {Kharel}, {Kilpatrick}, {Lanman}, {Lazda}, {Leung}, {Liu}, {Mas-Ribas}, {Masui}, {Mckinven}, {Mena-Parra}, {Miller}, {Nimmo}, {Pandhi}, {Pearlman}, {Pleunis}, {Prochaska}, {Rafiei-Ravandi}, {Sammons}, {Scholz}, {Shin}, {Smith}, {Stairs}, and {Swarali Shivraj}]{edf+24}
T.~{Eftekhari} et al.
\newblock \emph{arXiv e-prints}, art. arXiv:2410.23336, Oct. 2024.
\newblock \doi{10.48550/arXiv.2410.23336}.

\bibitem[{Eppel} et~al.(2025){Eppel}, {Krumpe}, {Limaye}, {Intrarat}, {Wongphechauxsorn}, {Cruces}, {Herrmann}, {Jankowski}, {Jaroenjittichai}, {Spitler}, and {Kadler}]{2025Eppel}
F.~{Eppel} et al.
\newblock \emph{\aap}, 695:\penalty0 L10, Mar. 2025.
\newblock \doi{10.1051/0004-6361/202453563}.

\bibitem[{Feng} et~al.(2022){Feng}, {Li}, {Yang}, {Zhang}, {Zhu}, {Zhang}, {Lu}, {Wang}, {Dai}, {Lynch}, {Yao}, {Jiang}, {Niu}, {Zhou}, {Xu}, {Miao}, {Niu}, {Meng}, {Qian}, {Tsai}, {Wang}, {Xue}, {Yue}, {Yuan}, {Zhang}, and {Zhang}]{2022Sci...375.1266F}
Y.~{Feng} et al.
\newblock \emph{Science}, 375\penalty0 (6586):\penalty0 1266--1270, Mar. 2022.
\newblock \doi{10.1126/science.abl7759}.

\bibitem[{Gajjar} et~al.(2018){Gajjar}, {Siemion}, {Price}, {Law}, {Michilli}, {Hessels}, {Chatterjee}, {Archibald}, {Bower}, {Brinkman}, {Burke-Spolaor}, {Cordes}, {Croft}, {Enriquez}, {Foster}, {Gizani}, {Hellbourg}, {Isaacson}, {Kaspi}, {Lazio}, {Lebofsky}, {Lynch}, {MacMahon}, {McLaughlin}, {Ransom}, {Scholz}, {Seymour}, {Spitler}, {Tendulkar}, {Werthimer}, and {Zhang}]{2018ApJ...863....2G}
V.~{Gajjar} et al.
\newblock \emph{\apj}, 863\penalty0 (1):\penalty0 2, Aug. 2018.
\newblock \doi{10.3847/1538-4357/aad005}.

\bibitem[{Gal-Yam}(2019)]{2019GalYam}
A.~{Gal-Yam}.
\newblock \emph{\araa}, 57:\penalty0 305--333, Aug. 2019.
\newblock \doi{10.1146/annurev-astro-081817-051819}.

\bibitem[{Gordon} et~al.(2023){Gordon}, {Fong}, {Kilpatrick}, {Eftekhari}, {Leja}, {Prochaska}, {Nugent}, {Bhandari}, {Blanchard}, {Caleb}, {Day}, {Deller}, {Dong}, {Glowacki}, {Gourdji}, {Mannings}, {Mahoney}, {Marnoch}, {Miller}, {Paterson}, {Rastinejad}, {Ryder}, {Sadler}, {Scott}, {Sears}, {Shannon}, {Simha}, {Stappers}, and {Tejos}]{2023ApJ...954...80G}
A.~C. {Gordon} et al.
\newblock \emph{\apj}, 954\penalty0 (1):\penalty0 80, Sept. 2023.
\newblock \doi{10.3847/1538-4357/ace5aa}.

\bibitem[{Gordon} et~al.(2025){Gordon}, {Fong}, {Deller}, {Marnoch}, {Lim}, {Peng}, {Bannister}, {Bera}, {Bhat}, {Dial}, {Dong}, {Eftekhari}, {Glowacki}, {Gourdji}, {Gupta}, {Jahns-Schindler}, {Jaini}, {Kilpatrick}, {Liu}, {Prochaska}, {Ryder}, {Shannon}, {Simha}, {Tejos}, {Wang}, and {Wang}]{2025ApJ...993..119G}
A.~C. {Gordon} et al.
\newblock \emph{\apj}, 993\penalty0 (1):\penalty0 119, Nov. 2025.
\newblock \doi{10.3847/1538-4357/ae0298}.

\bibitem[{Hallinan} et~al.(2019){Hallinan}, {Ravi}, {Weinreb}, {Kocz}, {Huang}, {Woody}, {Lamb}, {D'Addario}, {Catha}, {Law}, {Kulkarni}, {Phinney}, {Eastwood}, {Bouman}, {McLaughlin}, {Ransom}, {Siemens}, {Cordes}, {Lynch}, {Kaplan}, {Brazier}, {Bhatnagar}, {Myers}, {Walter}, and {Gaensler}]{2019BAAS...51g.255H}
G.~{Hallinan} et al.
\newblock In \emph{Bulletin of the American Astronomical Society}, volume~51, page 255, Sept. 2019.
\newblock \doi{10.48550/arXiv.1907.07648}.

\bibitem[{Hanmer} et~al.(2025){Hanmer}, {Pastor-Marazuela}, {Brink}, {Malesani}, {Stappers}, {Groot}, {Cooper}, {Tejos}, {Buckley}, {Barr}, {Bezuidenhout}, {Bloemen}, {Caleb}, {Driessen}, {Fender}, {Jankowski}, {Kramer}, {Pieterse}, {Rajwade}, {Tian}, {Vreeswijk}, {Wijnands}, and {Woudt}]{2025MNRAS.538.1800H}
K.~Y. {Hanmer} et al.
\newblock \emph{\mnras}, 538\penalty0 (3):\penalty0 1800--1815, Apr. 2025.
\newblock \doi{10.1093/mnras/staf289}.

\bibitem[{Hassall} et~al.(2012){Hassall}, {Stappers}, {Hessels}, {Kramer}, {Alexov}, {Anderson}, {Coenen}, {Karastergiou}, {Keane}, {Kondratiev}, {Lazaridis}, {van Leeuwen}, {Noutsos}, {Serylak}, {Sobey}, {Verbiest}, {Weltevrede}, {Zagkouris}, {Fender}, {Wijers}, {B{\"a}hren}, {Bell}, {Broderick}, {Corbel}, {Daw}, {Dhillon}, {Eisl{\"o}ffel}, {Falcke}, {Grie{\ss}meier}, {Jonker}, {Law}, {Markoff}, {Miller-Jones}, {Osten}, {Rol}, {Scaife}, {Scheers}, {Schellart}, {Spreeuw}, {Swinbank}, {ter Veen}, {Wise}, {Wijnands}, {Wucknitz}, {Zarka}, {Asgekar}, {Bell}, {Bentum}, {Bernardi}, {Best}, {Bonafede}, {Boonstra}, {Brentjens}, {Brouw}, {Br{\"u}ggen}, {Butcher}, {Ciardi}, {Garrett}, {Gerbers}, {Gunst}, {van Haarlem}, {Heald}, {Hoeft}, {Holties}, {de Jong}, {Koopmans}, {Kuniyoshi}, {Kuper}, {Loose}, {Maat}, {Masters}, {McKean}, {Meulman}, {Mevius}, {Munk}, {Noordam}, {Orr{\'u}}, {Paas}, {Pandey-Pommier}, {Pandey}, {Pizzo}, {Polatidis}, {Reich}, {R{\"o}ttgering}, {Sluman}, {Steinmetz}, {Sterks}, {Tagger}, {Tang},
  {Tasse}, {Vermeulen}, {van Weeren}, {Wijnholds}, and {Yatawatta}]{2012A&A...543A..66H}
T.~E. {Hassall} et al.
\newblock \emph{\aap}, 543:\penalty0 A66, July 2012.
\newblock \doi{10.1051/0004-6361/201218970}.

\bibitem[{Heintz} et~al.(2020){Heintz}, {Prochaska}, {Simha}, {Platts}, {Fong}, {Tejos}, {Ryder}, {Aggerwal}, {Bhandari}, {Day}, {Deller}, {Kilpatrick}, {Law}, {Macquart}, {Mannings}, {Marnoch}, {Sadler}, and {Shannon}]{2020ApJ...903..152H}
K.~E. {Heintz} et al.
\newblock \emph{\apj}, 903\penalty0 (2):\penalty0 152, Nov. 2020.
\newblock \doi{10.3847/1538-4357/abb6fb}.

\bibitem[{Hessels} et~al.(2019){Hessels}, {Spitler}, {Seymour}, {Cordes}, {Michilli}, {Lynch}, {Gourdji}, {Archibald}, {Bassa}, {Bower}, {Chatterjee}, {Connor}, {Crawford}, {Deneva}, {Gajjar}, {Kaspi}, {Keimpema}, {Law}, {Marcote}, {McLaughlin}, {Paragi}, {Petroff}, {Ransom}, {Scholz}, {Stappers}, and {Tendulkar}]{2019ApJ...876L..23H}
J.~W.~T. {Hessels} et al.
\newblock \emph{\apjl}, 876\penalty0 (2):\penalty0 L23, May 2019.
\newblock \doi{10.3847/2041-8213/ab13ae}.

\bibitem[{Hewish} et~al.(1968){Hewish}, {Bell}, {Pilkington}, {Scott}, and {Collins}]{JocelynBelllDiscoveryNature}
A.~{Hewish} et al.
\newblock \emph{\nat}, 217\penalty0 (5130):\penalty0 709--713, Feb. 1968.
\newblock \doi{10.1038/217709a0}.

\bibitem[{Houde} et~al.(2018){Houde}, {Mathews}, and {Rajabi}]{2018MNRAS.475..514H}
M.~{Houde}, A.~{Mathews}, and F.~{Rajabi}.
\newblock \emph{\mnras}, 475\penalty0 (1):\penalty0 514--522, Mar. 2018.
\newblock \doi{10.1093/mnras/stx3205}.

\bibitem[{Ivezi{\'c}} et~al.(2019){Ivezi{\'c}}, {Kahn}, {Tyson}, {Abel}, {Acosta}, {Allsman}, {Alonso}, {AlSayyad}, {Anderson}, {Andrew}, {Angel}, {Angeli}, {Ansari}, {Antilogus}, {Araujo}, {Armstrong}, {Arndt}, {Astier}, {Aubourg}, {Auza}, {Axelrod}, {Bard}, {Barr}, {Barrau}, {Bartlett}, {Bauer}, {Bauman}, {Baumont}, {Bechtol}, {Bechtol}, {Becker}, {Becla}, {Beldica}, {Bellavia}, {Bianco}, {Biswas}, {Blanc}, {Blazek}, {Blandford}, {Bloom}, {Bogart}, {Bond}, {Booth}, {Borgland}, {Borne}, {Bosch}, {Boutigny}, {Brackett}, {Bradshaw}, {Brandt}, {Brown}, {Bullock}, {Burchat}, {Burke}, {Cagnoli}, {Calabrese}, {Callahan}, {Callen}, {Carlin}, {Carlson}, {Chandrasekharan}, {Charles-Emerson}, {Chesley}, {Cheu}, {Chiang}, {Chiang}, {Chirino}, {Chow}, {Ciardi}, {Claver}, {Cohen-Tanugi}, {Cockrum}, {Coles}, {Connolly}, {Cook}, {Cooray}, {Covey}, {Cribbs}, {Cui}, {Cutri}, {Daly}, {Daniel}, {Daruich}, {Daubard}, {Daues}, {Dawson}, {Delgado}, {Dellapenna}, {de Peyster}, {de Val-Borro}, {Digel}, {Doherty}, {Dubois},
  {Dubois-Felsmann}, {Durech}, {Economou}, {Eifler}, {Eracleous}, {Emmons}, {Fausti Neto}, {Ferguson}, {Figueroa}, {Fisher-Levine}, {Focke}, {Foss}, {Frank}, {Freemon}, {Gangler}, {Gawiser}, {Geary}, {Gee}, {Geha}, {Gessner}, {Gibson}, {Gilmore}, {Glanzman}, {Glick}, {Goldina}, {Goldstein}, {Goodenow}, {Graham}, {Gressler}, {Gris}, {Guy}, {Guyonnet}, {Haller}, {Harris}, {Hascall}, {Haupt}, {Hernandez}, {Herrmann}, {Hileman}, {Hoblitt}, {Hodgson}, {Hogan}, {Howard}, {Huang}, {Huffer}, {Ingraham}, {Innes}, {Jacoby}, {Jain}, {Jammes}, {Jee}, {Jenness}, {Jernigan}, {Jevremovi{\'c}}, {Johns}, {Johnson}, {Johnson}, {Jones}, {Juramy-Gilles}, {Juri{\'c}}, {Kalirai}, {Kallivayalil}, {Kalmbach}, {Kantor}, {Karst}, {Kasliwal}, {Kelly}, {Kessler}, {Kinnison}, {Kirkby}, {Knox}, {Kotov}, {Krabbendam}, {Krughoff}, {Kub{\'a}nek}, {Kuczewski}, {Kulkarni}, {Ku}, {Kurita}, {Lage}, {Lambert}, {Lange}, {Langton}, {Le Guillou}, {Levine}, {Liang}, {Lim}, {Lintott}, {Long}, {Lopez}, {Lotz}, {Lupton}, {Lust}, {MacArthur}, {Mahabal},
  {Mandelbaum}, {Markiewicz}, {Marsh}, {Marshall}, {Marshall}, {May}, {McKercher}, {McQueen}, {Meyers}, {Migliore}, {Miller}, and {Mills}]{2019ApJ...873..111I}
{\v{Z}}.~{Ivezi{\'c}} et al.
\newblock \emph{\apj}, 873\penalty0 (2):\penalty0 111, Mar. 2019.
\newblock \doi{10.3847/1538-4357/ab042c}.

\bibitem[{Iwazaki}(2015)]{2015PhRvD..91b3008I}
A.~{Iwazaki}.
\newblock \emph{\prd}, 91\penalty0 (2):\penalty0 023008, Jan. 2015.
\newblock \doi{10.1103/PhysRevD.91.023008}.

\bibitem[{James} et~al.(2022){James}, {Prochaska}, {Macquart}, {North-Hickey}, {Bannister}, and {Dunning}]{2022MNRAS.510L..18J}
C.~W. {James} et al.
\newblock \emph{\mnras}, 510\penalty0 (1):\penalty0 L18--L23, Feb. 2022.
\newblock \doi{10.1093/mnrasl/slab117}.

\bibitem[{Jankowski} et~al.(2023){Jankowski}, {Bezuidenhout}, {Caleb}, {Driessen}, {Malenta}, {Morello}, {Rajwade}, {Sanidas}, {Stappers}, {Surnis}, {Barr}, {Chen}, {Kramer}, {Wu}, {Buchner}, {Serylak}, and {Prochaska}]{2023MNRAS.524.4275J}
F.~{Jankowski} et al.
\newblock \emph{\mnras}, 524\penalty0 (3):\penalty0 4275--4295, Sept. 2023.
\newblock \doi{10.1093/mnras/stad2041}.

\bibitem[{Kirsten} et~al.(2021){Kirsten}, {Snelders}, {Jenkins}, {Nimmo}, {van den Eijnden}, {Hessels}, {Gawro{\'n}ski}, and {Yang}]{2021NatAs...5..414K}
F.~{Kirsten} et al.
\newblock \emph{Nature Astronomy}, 5:\penalty0 414--422, Apr. 2021.
\newblock \doi{10.1038/s41550-020-01246-3}.

\bibitem[{Kirsten} et~al.(2022){Kirsten}, {Marcote}, {Nimmo}, {Hessels}, {Bhardwaj}, {Tendulkar}, {Keimpema}, {Yang}, {Snelders}, {Scholz}, {Pearlman}, {Law}, {Peters}, {Giroletti}, {Paragi}, {Bassa}, {Hewitt}, {Bach}, {Bezrukovs}, {Burgay}, {Buttaccio}, {Conway}, {Corongiu}, {Feiler}, {Forss{\'e}n}, {Gawro{\'n}ski}, {Karuppusamy}, {Kharinov}, {Lindqvist}, {Maccaferri}, {Melnikov}, {Ould-Boukattine}, {Possenti}, {Surcis}, {Wang}, {Yuan}, {Aggarwal}, {Anna-Thomas}, {Bower}, {Blaauw}, {Burke-Spolaor}, {Cassanelli}, {Clarke}, {Fonseca}, {Gaensler}, {Gopinath}, {Kaspi}, {Kassim}, {Lazio}, {Leung}, {Li}, {Lin}, {Masui}, {Mckinven}, {Michilli}, {Mikhailov}, {Ng}, {Orbidans}, {Pen}, {Petroff}, {Rahman}, {Ransom}, {Shin}, {Smith}, {Stairs}, and {Vlemmings}]{2022Natur.602..585K}
F.~{Kirsten} et al.
\newblock \emph{\nat}, 602\penalty0 (7898):\penalty0 585--589, Feb. 2022.
\newblock \doi{10.1038/s41586-021-04354-w}.

\bibitem[{Kremer} et~al.(2021){Kremer}, {Piro}, and {Li}]{2021ApJ...917L..11K}
K.~{Kremer}, A.~L. {Piro}, and D.~{Li}.
\newblock \emph{\apjl}, 917\penalty0 (1):\penalty0 L11, Aug. 2021.
\newblock \doi{10.3847/2041-8213/ac13a0}.

\bibitem[{Kumar} et~al.(2017){Kumar}, {Lu}, and {Bhattacharya}]{klb17}
P.~{Kumar}, W.~{Lu}, and M.~{Bhattacharya}.
\newblock \emph{MNRAS}, 468\penalty0 (3):\penalty0 2726--2739, Jul 2017.
\newblock \doi{10.1093/mnras/stx665}.

\bibitem[{Kumar} et~al.(2021){Kumar}, {Shannon}, {Flynn}, {Os{\l}owski}, {Bhandari}, {Day}, {Deller}, {Farah}, {Kaczmarek}, {Kerr}, {Phillips}, {Price}, {Qiu}, and {Thyagarajan}]{2021MNRAS.500.2525K}
P.~{Kumar} et al.
\newblock \emph{\mnras}, 500\penalty0 (2):\penalty0 2525--2531, Jan. 2021.
\newblock \doi{10.1093/mnras/staa3436}.

\bibitem[{Law} et~al.(2024){Law}, {Sharma}, {Ravi}, {Chen}, {Catha}, {Connor}, {Faber}, {Hallinan}, {Harnach}, {Hellbourg}, {Hobbs}, {Hodge}, {Hodges}, {Lamb}, {Rasmussen}, {Sherman}, {Shi}, {Simard}, {Squillace}, {Weinreb}, {Woody}, and {Yurk}]{2024ApJ...967...29L}
C.~J. {Law} et al.
\newblock \emph{\apj}, 967\penalty0 (1):\penalty0 29, May 2024.
\newblock \doi{10.3847/1538-4357/ad3736}.

\bibitem[{Leung} et~al.(2025){Leung}, {Simha}, {Medlock}, {Nagai}, {Masui}, {Kahinga}, {Lanman}, {Andrew}, {Bandura}, {Curtin}, {Gaensler}, {Gusinskaia}, {Joseph}, {Lazda}, {Mas-Ribas}, {Meyers}, {Nimmo}, {Pearlman}, {Prochaska}, {Sammons}, {Shin}, {Smith}, {Wang}, and {Chime/Frb Collaboration}]{2025ApJ...991L..25L}
C.~{Leung} et al.
\newblock \emph{\apjl}, 991\penalty0 (1):\penalty0 L25, Sept. 2025.
\newblock \doi{10.3847/2041-8213/ae044d}.

\bibitem[{Li} et~al.(2021{\natexlab{a}}){Li}, {Lin}, {Xiong}, {Ge}, {Li}, {Li}, {Lu}, {Zhang}, {Tuo}, {Nang}, {Zhang}, {Xiao}, {Chen}, {Song}, {Xu}, {Liu}, {Jia}, {Cao}, {Qu}, {Zhang}, {Gu}, {Liao}, {Zhao}, {Tan}, {Nie}, {Zhao}, {Zheng}, {Zheng}, {Luo}, {Cai}, {Li}, {Xue}, {Bu}, {Chang}, {Chen}, {Chen}, {Chen}, {Chen}, {Chen}, {Cui}, {Cui}, {Deng}, {Dong}, {Du}, {Fu}, {Gao}, {Gao}, {Gao}, {Gu}, {Guan}, {Guo}, {Han}, {Huang}, {Huo}, {Jiang}, {Jiang}, {Jin}, {Jin}, {Kong}, {Li}, {Li}, {Li}, {Li}, {Li}, {Li}, {Li}, {Liang}, {Liu}, {Liu}, {Liu}, {Liu}, {Liu}, {Lu}, {Lu}, {Luo}, {Ma}, {Meng}, {Ou}, {Sai}, {Shang}, {Song}, {Sun}, {Tao}, {Wang}, {Wang}, {Wang}, {Wang}, {Wang}, {Wen}, {Wu}, {Wu}, {Wu}, {Xiao}, {Xu}, {Yang}, {Yang}, {Yang}, {Yang}, {Yi}, {Yin}, {You}, {Zhang}, {Zhang}, {Zhang}, {Zhang}, {Zhang}, {Zhang}, {Zhang}, {Zhang}, {Zhang}, {Zhang}, {Zhang}, {Zhang}, {Zhang}, {Zhang}, {Zhang}, {Zhang}, {Zhou}, {Zhou}, {Zhu}, {Zhu}, and {Zhuang}]{2021NatureInsightSGR}
C.~K. {Li} et al.
\newblock \emph{Nature Astronomy}, 5:\penalty0 378--384, Apr. 2021{\natexlab{a}}.
\newblock \doi{10.1038/s41550-021-01302-6}.

\bibitem[{Li} et~al.(2021{\natexlab{b}}){Li}, {Wang}, {Zhu}, {Zhang}, {Zhang}, {Duan}, {Zhang}, {Feng}, {Tang}, {Chatterjee}, {Cordes}, {Cruces}, {Dai}, {Gajjar}, {Hobbs}, {Jin}, {Kramer}, {Lorimer}, {Miao}, {Niu}, {Niu}, {Pan}, {Qian}, {Spitler}, {Werthimer}, {Zhang}, {Wang}, {Xie}, {Yue}, {Zhang}, {Zhi}, and {Zhu}]{2021Natur.598..267L}
D.~{Li} et al.
\newblock \emph{\nat}, 598\penalty0 (7880):\penalty0 267--271, Oct. 2021{\natexlab{b}}.
\newblock \doi{10.1038/s41586-021-03878-5}.

\bibitem[{Limaye} et~al.(2025){Limaye}, {Spitler}, {Manaswini}, {Ben{\'a}{\v{c}}ek}, {Eppel}, {Kadler}, {Nicotera}, and {Wongphechauxsorn}]{2025arXiv251008367L}
P.~{Limaye} et al.
\newblock \emph{arXiv e-prints}, art. arXiv:2510.08367, Oct. 2025.
\newblock \doi{10.48550/arXiv.2510.08367}.

\bibitem[{Lin} et~al.(2022){Lin}, {Lin}, {Li}, {Tseng}, {Jiang}, {Wang}, {Cheng}, {Pen}, {Chen}, {Chen}, {Chen}, {Goto}, {Hashimoto}, {Hwang}, {King}, {Kubo}, {Kuo}, {Mills}, {Nam}, {Oshiro}, {Shen}, {Tseng}, {Wang}, {Wu}, {Bower}, {Chang}, {Chen}, {Chen}, {Chiang}, {Fedynitch}, {Gusinskaia}, {Ho}, {Hsiao}, {Hu}, {Huang}, {J{\'a}uregui Garc{\'\i}a}, {Kim}, {Kuo}, {Ling}, {On}, {Peterson}, {R. Raquel}, {Su}, {Uno}, {Wu}, {Yamasaki}, and {Zhu}]{2022PASP..134i4106L}
H.-H. {Lin} et al.
\newblock \emph{\pasp}, 134\penalty0 (1039):\penalty0 094106, Sept. 2022.
\newblock \doi{10.1088/1538-3873/ac8f71}.

\bibitem[{Lorimer} et~al.(2007){Lorimer}, {Bailes}, {McLaughlin}, {Narkevic}, and {Crawford}]{LorimerBurst}
D.~R. {Lorimer} et al.
\newblock \emph{Science}, 318\penalty0 (5851):\penalty0 777, Nov. 2007.
\newblock \doi{10.1126/science.1147532}.

\bibitem[Lu and Kumar(2018)]{lu2018radiation}
W.~Lu and P.~Kumar.
\newblock \emph{Monthly Notices of the Royal Astronomical Society}, 477\penalty0 (2):\penalty0 2470--2493, 2018.

\bibitem[{Lyubarsky}(2014)]{2014MNRAS.442L...9L}
Y.~{Lyubarsky}.
\newblock \emph{\mnras}, 442:\penalty0 L9--L13, July 2014.
\newblock \doi{10.1093/mnrasl/slu046}.

\bibitem[{Mannings} et~al.(2021){Mannings}, {Fong}, {Simha}, {Prochaska}, {Rafelski}, {Kilpatrick}, {Tejos}, {Heintz}, {Bannister}, {Bhandari}, {Day}, {Deller}, {Ryder}, {Shannon}, and {Tendulkar}]{mfs+21}
A.~G. {Mannings} et al.
\newblock \emph{ApJ}, 917\penalty0 (2):\penalty0 75, Aug. 2021.
\newblock \doi{10.3847/1538-4357/abff56}.

\bibitem[{Marcote} et~al.(2017){Marcote}, {Paragi}, {Hessels}, {Keimpema}, {van Langevelde}, {Huang}, {Bassa}, {Bogdanov}, {Bower}, {Burke-Spolaor}, {Butler}, {Campbell}, {Chatterjee}, {Cordes}, {Demorest}, {Garrett}, {Ghosh}, {Kaspi}, {Law}, {Lazio}, {McLaughlin}, {Ransom}, {Salter}, {Scholz}, {Seymour}, {Siemion}, {Spitler}, {Tendulkar}, and {Wharton}]{2017ApJ...834L...8M}
B.~{Marcote} et al.
\newblock \emph{\apjl}, 834\penalty0 (2):\penalty0 L8, Jan. 2017.
\newblock \doi{10.3847/2041-8213/834/2/L8}.

\bibitem[{Marcote} et~al.(2020){Marcote}, {Nimmo}, {Hessels}, {Tendulkar}, {Bassa}, {Paragi}, {Keimpema}, {Bhardwaj}, {Karuppusamy}, {Kaspi}, {Law}, {Michilli}, {Aggarwal}, {Andersen}, {Archibald}, {Bandura}, {Bower}, {Boyle}, {Brar}, {Burke-Spolaor}, {Butler}, {Cassanelli}, {Chawla}, {Demorest}, {Dobbs}, {Fonseca}, {Giri}, {Good}, {Gourdji}, {Josephy}, {Kirichenko}, {Kirsten}, {Landecker}, {Lang}, {Lazio}, {Li}, {Lin}, {Linford}, {Masui}, {Mena-Parra}, {Naidu}, {Ng}, {Patel}, {Pen}, {Pleunis}, {Rafiei-Ravandi}, {Rahman}, {Renard}, {Scholz}, {Siegel}, {Smith}, {Stairs}, {Vanderlinde}, and {Zwaniga}]{2020Natur.577..190M}
B.~{Marcote} et al.
\newblock \emph{\nat}, 577\penalty0 (7789):\penalty0 190--194, Jan. 2020.
\newblock \doi{10.1038/s41586-019-1866-z}.

\bibitem[Margalit and Metzger(2018)]{Margalit_2018}
B.~Margalit and B.~D. Metzger.
\newblock \emph{The Astrophysical Journal Letters}, 868\penalty0 (1):\penalty0 L4, Nov. 2018.
\newblock ISSN 2041-8213.
\newblock \doi{10.3847/2041-8213/aaedad}.
\newblock URL \url{http://dx.doi.org/10.3847/2041-8213/aaedad}.

\bibitem[{Mckinven} et~al.(2023){Mckinven}, {Gaensler}, {Michilli}, {Masui}, {Kaspi}, {Bhardwaj}, {Cassanelli}, {Chawla}, {Dong}, {Fonseca}, {Leung}, {Li}, {Ng}, {Patel}, {Petroff}, {Pearlman}, {Pleunis}, {Rafiei-Ravandi}, {Rahman}, {Sand}, {Shin}, {Scholz}, {Stairs}, {Smith}, {Su}, and {Tendulkar}]{2023ApJ...950...12M}
R.~{Mckinven} et al.
\newblock \emph{\apj}, 950\penalty0 (1):\penalty0 12, June 2023.
\newblock \doi{10.3847/1538-4357/acc65f}.

\bibitem[{Mckinven} et~al.(2025){Mckinven}, {Bhardwaj}, {Eftekhari}, {Kilpatrick}, {Kirichenko}, {Pal}, {Cook}, {Gaensler}, {Giri}, {Kaspi}, {Michilli}, {Nimmo}, {Pearlman}, {Pleunis}, {Sand}, {Stairs}, {Andersen}, {Andrew}, {Bandura}, {Brar}, {Cassanelli}, {Chatterjee}, {Curtin}, {Dong}, {Eadie}, {Fonseca}, {Ibik}, {Kaczmarek}, {Kharel}, {Lazda}, {Leung}, {Li}, {Main}, {Masui}, {Mena-Parra}, {Ng}, {Pandhi}, {Patil}, {Prochaska}, {Rafiei-Ravandi}, {Scholz}, {Shah}, {Shin}, and {Smith}]{2025Natur.637...43M}
R.~{Mckinven} et al.
\newblock \emph{\nat}, 637\penalty0 (8044):\penalty0 43--47, Jan. 2025.
\newblock \doi{10.1038/s41586-024-08184-4}.

\bibitem[{Mereghetti} et~al.(2020){Mereghetti}, {Savchenko}, {Ferrigno}, {G{\"o}tz}, {Rigoselli}, {Tiengo}, {Bazzano}, {Bozzo}, {Coleiro}, {Courvoisier}, {Doyle}, {Goldwurm}, {Hanlon}, {Jourdain}, {von Kienlin}, {Lutovinov}, {Martin-Carrillo}, {Molkov}, {Natalucci}, {Onori}, {Panessa}, {Rodi}, {Rodriguez}, {S{\'a}nchez-Fern{\'a}ndez}, {Sunyaev}, and {Ubertini}]{msf+20}
S.~{Mereghetti} et al.
\newblock \emph{ApJL}, 898\penalty0 (2):\penalty0 L29, Aug. 2020.
\newblock \doi{10.3847/2041-8213/aba2cf}.

\bibitem[{Metzger} et~al.(2017){Metzger}, {Berger}, and {Margalit}]{2017ApJ...841...14M}
B.~D. {Metzger}, E.~{Berger}, and B.~{Margalit}.
\newblock \emph{\apj}, 841\penalty0 (1):\penalty0 14, May 2017.
\newblock \doi{10.3847/1538-4357/aa633d}.

\bibitem[{Metzger} et~al.(2019){Metzger}, {Margalit}, and {Sironi}]{mms19}
B.~D. {Metzger}, B.~{Margalit}, and L.~{Sironi}.
\newblock \emph{MNRAS}, 485\penalty0 (3):\penalty0 4091--4106, May 2019.
\newblock \doi{10.1093/mnras/stz700}.

\bibitem[Michilli et~al.(2018)Michilli, Seymour, Hessels, Spitler, Gajjar, Archibald, Bower, Chatterjee, Cordes, Gourdji, et~al.]{michilli2018extreme}
D.~Michilli et al.
\newblock \emph{Nature}, 553\penalty0 (7687):\penalty0 182, 2018.

\bibitem[{Mingarelli} et~al.(2015){Mingarelli}, {Levin}, and {Lazio}]{Mingarelli2015}
C.~M.~F. {Mingarelli}, J.~{Levin}, and T.~J.~W. {Lazio}.
\newblock \emph{ApJL}, 814\penalty0 (2):\penalty0 L20, Dec. 2015.
\newblock \doi{10.1088/2041-8205/814/2/L20}.

\bibitem[Moroianu et~al.(2025)Moroianu, Bhandari, Drout, Hessels, Hewitt, Kirsten, Marcote, Pleunis, Snelders, Sridhar, Bach, Bempong-Manful, Bezrukovs, Blaauw, Bray, Buttaccio, Chatterjee, Corongiu, Feiler, Gaensler, Gawroński, Giroletti, Ibik, Karuppusamy, Lazda, Leung, Lindqvist, Masui, Michilli, Nimmo, Ould-Boukattine, Pandhi, Paragi, Pearlman, Puchalska, Scholz, Shin, Sluman, Trudu, Williams-Baldwin, and Yang]{moroianu2025}
A.~M. Moroianu et al.
\newblock A milliarcsecond localization associates frb 20190417a with a compact, luminous persistent radio source and an extreme magneto-ionic environment, 2025.
\newblock URL \url{https://arxiv.org/abs/2509.05174}.

\bibitem[{Most} and {Philippov}(2020)]{MostPhilippov2020}
E.~R. {Most} and A.~A. {Philippov}.
\newblock \emph{ApJL}, 893\penalty0 (1):\penalty0 L6, Apr. 2020.
\newblock \doi{10.3847/2041-8213/ab8196}.

\bibitem[{Most} and {Philippov}(2022)]{2022MostPhilippov}
E.~R. {Most} and A.~A. {Philippov}.
\newblock \emph{MNRAS}, 515\penalty0 (2):\penalty0 2710--2724, Sept. 2022.
\newblock \doi{10.1093/mnras/stac1909}.

\bibitem[{Most} and {Philippov}(2023)]{2023MostPhilippov}
E.~R. {Most} and A.~A. {Philippov}.
\newblock \emph{\prl}, 130\penalty0 (24):\penalty0 245201, June 2023.
\newblock \doi{10.1103/PhysRevLett.130.245201}.

\bibitem[{Mottez} and {Zarka}(2014)]{2014A&A...569A..86M}
F.~{Mottez} and P.~{Zarka}.
\newblock \emph{\aap}, 569:\penalty0 A86, Sept. 2014.
\newblock \doi{10.1051/0004-6361/201424104}.

\bibitem[{Nimmo} et~al.(2022){Nimmo}, {Hessels}, {Kirsten}, {Keimpema}, {Cordes}, {Snelders}, {Hewitt}, {Karuppusamy}, {Archibald}, {Bezrukovs}, {Bhardwaj}, {Blaauw}, {Buttaccio}, {Cassanelli}, {Conway}, {Corongiu}, {Feiler}, {Fonseca}, {Forss{\'e}n}, {Gawro{\'n}ski}, {Giroletti}, {Kharinov}, {Leung}, {Lindqvist}, {Maccaferri}, {Marcote}, {Masui}, {Mckinven}, {Melnikov}, {Michilli}, {Mikhailov}, {Ng}, {Orbidans}, {Ould-Boukattine}, {Paragi}, {Pearlman}, {Petroff}, {Rahman}, {Scholz}, {Shin}, {Smith}, {Stairs}, {Surcis}, {Tendulkar}, {Vlemmings}, {Wang}, {Yang}, and {Yuan}]{2022NatAs...6..393N}
K.~{Nimmo} et al.
\newblock \emph{Nature Astronomy}, 6:\penalty0 393--401, Feb. 2022.
\newblock \doi{10.1038/s41550-021-01569-9}.

\bibitem[{Nimmo} et~al.(2025){Nimmo}, {Pleunis}, {Beniamini}, {Kumar}, {Lanman}, {Li}, {Main}, {Sammons}, {Andrew}, {Bhardwaj}, {Chatterjee}, {Curtin}, {Fonseca}, {Gaensler}, {Joseph}, {Kader}, {Kaspi}, {Lazda}, {Leung}, {Masui}, {Mckinven}, {Michilli}, {Pandhi}, {Pearlman}, {Rafiei-Ravandi}, {Sand}, {Shin}, {Smith}, and {Stairs}]{2025Natur.637...48N}
K.~{Nimmo} et al.
\newblock \emph{\nat}, 637\penalty0 (8044):\penalty0 48--51, Jan. 2025.
\newblock \doi{10.1038/s41586-024-08297-w}.

\bibitem[{Niu} et~al.(2022){Niu}, {Aggarwal}, {Li}, {Zhang}, {Chatterjee}, {Tsai}, {Yu}, {Law}, {Burke-Spolaor}, {Cordes}, {Zhang}, {Ocker}, {Yao}, {Wang}, {Feng}, {Niino}, {Bochenek}, {Cruces}, {Connor}, {Jiang}, {Dai}, {Luo}, {Li}, {Miao}, {Niu}, {Anna-Thomas}, {Sydnor}, {Stern}, {Wang}, {Yuan}, {Yue}, {Zhou}, {Yan}, {Zhu}, and {Zhang}]{2022Natur.606..873N}
C.-H. {Niu} et al.
\newblock \emph{\nat}, 606\penalty0 (7916):\penalty0 873--877, June 2022.
\newblock \doi{10.1038/s41586-022-04755-5}.

\bibitem[{Ocker} et~al.(2022{\natexlab{a}}){Ocker}, {Cordes}, {Chatterjee}, and {Gorsuch}]{2022ApJ...934...71O}
S.~K. {Ocker}, J.~M. {Cordes}, S.~{Chatterjee}, and M.~R. {Gorsuch}.
\newblock \emph{\apj}, 934\penalty0 (1):\penalty0 71, July 2022{\natexlab{a}}.
\newblock \doi{10.3847/1538-4357/ac75ba}.

\bibitem[{Ocker} et~al.(2022{\natexlab{b}}){Ocker}, {Cordes}, {Chatterjee}, {Niu}, {Li}, {McKee}, {Law}, {Tsai}, {Anna-Thomas}, {Yao}, and {Cruces}]{2022ApJ...931...87O}
S.~K. {Ocker} et al.
\newblock \emph{\apj}, 931\penalty0 (2):\penalty0 87, June 2022{\natexlab{b}}.
\newblock \doi{10.3847/1538-4357/ac6504}.

\bibitem[{Ould-Boukattine} et~al.(2025){Ould-Boukattine}, {Cooper}, {Hessels}, {Hewitt}, {Ocker}, {Moroianu}, {Nimmo}, {Snelders}, {Cognard}, {Dijkema}, {Fine}, {Gawro{\'n}ski}, {Herrmann}, {Huang}, {Kirsten}, {Pleunis}, {Puchalska}, {Ranguin}, and {Telkamp}]{2025arXiv250916374O}
O.~S. {Ould-Boukattine} et al.
\newblock \emph{arXiv e-prints}, art. arXiv:2509.16374, Sept. 2025.
\newblock \doi{10.48550/arXiv.2509.16374}.

\bibitem[{Paine} et~al.(2024){Paine}, {Hawkins}, {Lorimer}, {Stanley}, {Kania}, {Crawford}, and {Fairfield}]{2024MNRAS.528.6340P}
S.~{Paine} et al.
\newblock \emph{\mnras}, 528\penalty0 (4):\penalty0 6340--6346, Mar. 2024.
\newblock \doi{10.1093/mnras/stae344}.

\bibitem[Pandhi et~al.(2026)Pandhi, Nimmo, Andrew, Brar, Chatterjee, Cook, Curtin, Gaensler, Gawronski, Hessels, Kaspi, Khan, Kirsten, Lazda, Leung, Main, Masui, Mckinven, Michilli, Ng, Ould-Boukattine, Pearlman, Pleunis, Pollak, Pradeep E.~T., Puchalska, Sammons, Scholz, Shah, Shin, Siegel, and Smith]{Pandhi_2026}
A.~Pandhi et al.
\newblock \emph{The Astrophysical Journal Letters}, 1000\penalty0 (2):\penalty0 L53, mar 2026.
\newblock \doi{10.3847/2041-8213/ae52f8}.
\newblock URL \url{https://doi.org/10.3847/2041-8213/ae52f8}.

\bibitem[{Pastor-Marazuela} et~al.(2021){Pastor-Marazuela}, {Connor}, {van Leeuwen}, {Maan}, {ter Veen}, {Bilous}, {Oostrum}, {Petroff}, {Straal}, {Vohl}, {Attema}, {Boersma}, {Kooistra}, {van der Schuur}, {Sclocco}, {Smits}, {Adams}, {Adebahr}, {de Blok}, {Coolen}, {Damstra}, {D{\'e}nes}, {Hess}, {van der Hulst}, {Hut}, {Ivashina}, {Kutkin}, {Loose}, {Lucero}, {Mika}, {Moss}, {Mulder}, {Norden}, {Oosterloo}, {Orr{\'u}}, {Ruiter}, and {Wijnholds}]{2021Natur.596..505P}
I.~{Pastor-Marazuela} et al.
\newblock \emph{\nat}, 596\penalty0 (7873):\penalty0 505--508, Aug. 2021.
\newblock \doi{10.1038/s41586-021-03724-8}.

\bibitem[{Pastor-Marazuela} et~al.(2025){Pastor-Marazuela}, {Gordon}, {Stappers}, {Khrykin}, {Tejos}, {Rajwade}, {Caleb}, {Surnis}, {Driessen}, {Simha}, {Tian}, {Prochaska}, {Barr}, {Fong}, {Jankowski}, {Kahinga}, {Kilpatrick}, {Kramer}, and {Mas-Ribas}]{2025arXiv250705982P}
I.~{Pastor-Marazuela} et al.
\newblock \emph{arXiv e-prints}, art. arXiv:2507.05982, July 2025.
\newblock \doi{10.48550/arXiv.2507.05982}.

\bibitem[{Pearlman} et~al.(2025){Pearlman}, {Scholz}, {Bethapudi}, {Hessels}, {Kaspi}, {Kirsten}, {Nimmo}, {Spitler}, {Fonseca}, {Meyers}, {Stairs}, {Tan}, {Bhardwaj}, {Chatterjee}, {Cook}, {Curtin}, {Dong}, {Eftekhari}, {Gaensler}, {G{\"u}ver}, {Kaczmarek}, {Leung}, {Masui}, {Michilli}, {Prince}, {Sand}, {Shin}, {Smith}, and {Tendulkar}]{2025NatAs...9..111P}
A.~B. {Pearlman} et al.
\newblock \emph{Nature Astronomy}, 9:\penalty0 111--127, Jan. 2025.
\newblock \doi{10.1038/s41550-024-02386-6}.

\bibitem[{Pelliciari} et~al.(2023){Pelliciari}, {Bernardi}, {Pilia}, {Naldi}, {Pupillo}, {Trudu}, {Addis}, {Bianchi}, {Bortolotti}, {Dallacasa}, {Lulli}, {Maccaferri}, {Magro}, {Mattana}, {Perini}, {Roma}, {Schiaffino}, {Setti}, {Tavani}, {Verrecchia}, and {Casentini}]{2023A&A...674A.223P}
D.~{Pelliciari} et al.
\newblock \emph{\aap}, 674:\penalty0 A223, June 2023.
\newblock \doi{10.1051/0004-6361/202346307}.

\bibitem[{Platts} et~al.(2019){Platts}, {Weltman}, {Walters}, {Tendulkar}, {Gordin}, and {Kandhai}]{2019PhR...821....1P}
E.~{Platts} et al.
\newblock \emph{\physrep}, 821:\penalty0 1--27, Aug. 2019.
\newblock \doi{10.1016/j.physrep.2019.06.003}.

\bibitem[{Pleunis} et~al.(2021{\natexlab{a}}){Pleunis}, {Good}, {Kaspi}, {Mckinven}, {Ransom}, {Scholz}, {Bandura}, {Bhardwaj}, {Boyle}, {Brar}, {Cassanelli}, {Chawla}, {(Adam) Dong}, {Fonseca}, {Gaensler}, {Josephy}, {Kaczmarek}, {Leung}, {Lin}, {Masui}, {Mena-Parra}, {Michilli}, {Ng}, {Patel}, {Rafiei-Ravandi}, {Rahman}, {Sanghavi}, {Shin}, {Smith}, {Stairs}, and {Tendulkar}]{pgk+21}
Z.~{Pleunis} et al.
\newblock \emph{ApJ}, 923\penalty0 (1):\penalty0 1, Dec. 2021{\natexlab{a}}.
\newblock \doi{10.3847/1538-4357/ac33ac}.

\bibitem[{Pleunis} et~al.(2021{\natexlab{b}}){Pleunis}, {Michilli}, {Bassa}, {Hessels}, {Naidu}, {Andersen}, {Chawla}, {Fonseca}, {Gopinath}, {Kaspi}, {Kondratiev}, {Li}, {Bhardwaj}, {Boyle}, {Brar}, {Cassanelli}, {Gupta}, {Josephy}, {Karuppusamy}, {Keimpema}, {Kirsten}, {Leung}, {Marcote}, {Masui}, {Mckinven}, {Meyers}, {Ng}, {Nimmo}, {Paragi}, {Rahman}, {Scholz}, {Shin}, {Smith}, {Stairs}, and {Tendulkar}]{2021ApJ...911L...3P}
Z.~{Pleunis} et al.
\newblock \emph{\apjl}, 911\penalty0 (1):\penalty0 L3, Apr. 2021{\natexlab{b}}.
\newblock \doi{10.3847/2041-8213/abec72}.

\bibitem[{Rajwade} et~al.(2022){Rajwade}, {Bezuidenhout}, {Caleb}, {Driessen}, {Jankowski}, {Malenta}, {Morello}, {Sanidas}, {Stappers}, {Surnis}, {Barr}, {Chen}, {Kramer}, {Wu}, {Buchner}, {Serylak}, {Combes}, {Fong}, {Gupta}, {Jagannathan}, {Kilpatrick}, {Krogager}, {Noterdaeme}, {N{\'u}nẽz}, {Prochaska}, {Srianand}, and {Tejos}]{2022MNRAS.514.1961R}
K.~M. {Rajwade} et al.
\newblock \emph{\mnras}, 514\penalty0 (2):\penalty0 1961--1974, Aug. 2022.
\newblock \doi{10.1093/mnras/stac1450}.

\bibitem[{Ravi}(2019)]{2019NatAs...3..928R}
V.~{Ravi}.
\newblock \emph{Nature Astronomy}, 3:\penalty0 928--931, July 2019.
\newblock \doi{10.1038/s41550-019-0831-y}.

\bibitem[{Rhodes} et~al.(2023){Rhodes}, {Caleb}, {Stappers}, {Andersson}, {Bezuidenhout}, {Driessen}, and {Heywood}]{2023MNRAS.525.3626R}
L.~{Rhodes} et al.
\newblock \emph{\mnras}, 525\penalty0 (3):\penalty0 3626--3632, Nov. 2023.
\newblock \doi{10.1093/mnras/stad2438}.

\bibitem[{Ridnaia} et~al.(2021){Ridnaia}, {Svinkin}, {Frederiks}, {Bykov}, {Popov}, {Aptekar}, {Golenetskii}, {Lysenko}, {Tsvetkova}, {Ulanov}, and {Cline}]{rsf+21}
A.~{Ridnaia} et al.
\newblock \emph{Nature Astronomy}, 5:\penalty0 372--377, Apr. 2021.
\newblock \doi{10.1038/s41550-020-01265-0}.

\bibitem[{Ryder} et~al.(2023){Ryder}, {Bannister}, {Bhandari}, {Deller}, {Ekers}, {Glowacki}, {Gordon}, {Gourdji}, {James}, {Kilpatrick}, {Lu}, {Marnoch}, {Moss}, {Prochaska}, {Qiu}, {Sadler}, {Simha}, {Sammons}, {Scott}, {Tejos}, and {Shannon}]{2023Sci...382..294R}
S.~D. {Ryder} et al.
\newblock \emph{Science}, 382\penalty0 (6668):\penalty0 294--299, Oct. 2023.
\newblock \doi{10.1126/science.adf2678}.

\bibitem[{Sand} et~al.(2024){Sand}, {Curtin}, {Michilli}, {Kaspi}, {Fonseca}, {Nimmo}, {Pleunis}, {Shin}, {Bhardwaj}, {Brar}, {Dobbs}, {Eadie}, {Gaensler}, {Joseph}, {Leung}, {Main}, {Masui}, {Mckinven}, {Pandhi}, {Pearlman}, {Rafiei-Ravandi}, {Sammons}, {Smith}, and {Stairs}]{scm+24}
K.~R. {Sand} et al.
\newblock \emph{arXiv e-prints}, art. arXiv:2408.13215, Aug. 2024.
\newblock \doi{10.48550/arXiv.2408.13215}.

\bibitem[{Scholz} et~al.(2020){Scholz}, {Cook}, {Cruces}, {Hessels}, {Kaspi}, {Majid}, {Naidu}, {Pearlman}, {Spitler}, {Bandura}, {Bhardwaj}, {Cassanelli}, {Chawla}, {Gaensler}, {Good}, {Josephy}, {Karuppusamy}, {Keimpema}, {Kirichenko}, {Kirsten}, {Kocz}, {Leung}, {Marcote}, {Masui}, {Mena-Parra}, {Merryfield}, {Michilli}, {Naudet}, {Nimmo}, {Pleunis}, {Prince}, {Rafiei-Ravandi}, {Rahman}, {Shin}, {Smith}, {Stairs}, {Tendulkar}, and {Vanderlinde}]{scholz2020}
P.~{Scholz} et al.
\newblock \emph{\apj}, 901\penalty0 (2):\penalty0 165, Oct. 2020.
\newblock \doi{10.3847/1538-4357/abb1a8}.

\bibitem[{Shah} et~al.(2024){Shah}, {Shin}, {Leung}, {Fong}, {Eftekhari}, {Amiri}, {Andersen}, {Andrew}, {Bhardwaj}, {Brar}, {Cassanelli}, {Chatterjee}, {Curtin}, {Dobbs}, {Dong}, {Dong}, {Fonseca}, {Gaensler}, {Halpern}, {Hessels}, {Ibik}, {Jain}, {Joseph}, {Kaczmarek}, {Kahinga}, {Kaspi}, {Kharel}, {Landecker}, {Lanman}, {Lazda}, {Main}, {Mas-Ribas}, {Masui}, {Mckinven}, {Mena-Parra}, {Meyers}, {Michilli}, {Nimmo}, {Pandhi}, {Patil}, {Pearlman}, {Pleunis}, {Prochaska}, {Rafiei-Ravandi}, {Sammons}, {Sand}, {Scholz}, {Smith}, and {Stairs}]{ssl+24}
V.~{Shah} et al.
\newblock \emph{arXiv e-prints}, art. arXiv:2410.23374, Oct. 2024.
\newblock \doi{10.48550/arXiv.2410.23374}.

\bibitem[Shannon et~al.(2025)Shannon, Bannister, Bera, Bhandari, Day, Deller, Dial, Dobie, Ekers, Fong, Glowacki, Gordon, Gourdji, Jaini, James, Kumar, Mahony, Marnoch, Muller, Prochaska, Qiu, Ryder, Sadler, Scott, Tejos, Uttarkar, and Wang]{Shannon_2025}
R.~M. Shannon et al.
\newblock \emph{Publications of the Astronomical Society of Australia}, 42, 2025.
\newblock ISSN 1448-6083.
\newblock \doi{10.1017/pasa.2025.8}.
\newblock URL \url{http://dx.doi.org/10.1017/pasa.2025.8}.

\bibitem[{Snelders} et~al.(2023){Snelders}, {Nimmo}, {Hessels}, {Bensellam}, {Zwaan}, {Chawla}, {Ould-Boukattine}, {Kirsten}, {Faber}, and {Gajjar}]{2023NatAs...7.1486S}
M.~P. {Snelders} et al.
\newblock \emph{Nature Astronomy}, 7:\penalty0 1486--1496, Dec. 2023.
\newblock \doi{10.1038/s41550-023-02101-x}.

\bibitem[{Spitler} et~al.(2014){Spitler}, {Cordes}, {Hessels}, {Lorimer}, {McLaughlin}, {Chatterjee}, {Crawford}, {Deneva}, {Kaspi}, {Wharton}, {Allen}, {Bogdanov}, {Brazier}, {Camilo}, {Freire}, {Jenet}, {Karako-Argaman}, {Knispel}, {Lazarus}, {Lee}, {van Leeuwen}, {Lynch}, {Ransom}, {Scholz}, {Siemens}, {Stairs}, {Stovall}, {Swiggum}, {Venkataraman}, {Zhu}, {Aulbert}, and {Fehrmann}]{2014ApJ...790..101S}
L.~G. {Spitler} et al.
\newblock \emph{\apj}, 790\penalty0 (2):\penalty0 101, Aug. 2014.
\newblock \doi{10.1088/0004-637X/790/2/101}.

\bibitem[{Spitler} et~al.(2016){Spitler}, {Scholz}, {Hessels}, {Bogdanov}, {Brazier}, {Camilo}, {Chatterjee}, {Cordes}, {Crawford}, {Deneva}, {Ferdman}, {Freire}, {Kaspi}, {Lazarus}, {Lynch}, {Madsen}, {McLaughlin}, {Patel}, {Ransom}, {Seymour}, {Stairs}, {Stappers}, {van Leeuwen}, and {Zhu}]{2016Natur.531..202S}
L.~G. {Spitler} et al.
\newblock \emph{\nat}, 531\penalty0 (7593):\penalty0 202--205, Mar. 2016.
\newblock \doi{10.1038/nature17168}.

\bibitem[{Sridhar} et~al.(2021){Sridhar}, {Metzger}, {Beniamini}, {Margalit}, {Renzo}, {Sironi}, and {Kovlakas}]{2021ApJ...917...13S}
N.~{Sridhar} et al.
\newblock \emph{\apj}, 917\penalty0 (1):\penalty0 13, Aug. 2021.
\newblock \doi{10.3847/1538-4357/ac0140}.

\bibitem[{Tendulkar} et~al.(2021){Tendulkar}, {Gil de Paz}, {Kirichenko}, {Hessels}, {Bhardwaj}, {{\'A}vila}, {Bassa}, {Chawla}, {Fonseca}, {Kaspi}, {Keimpema}, {Kirsten}, {Lazio}, {Marcote}, {Masui}, {Nimmo}, {Paragi}, {Rahman}, {Pay{\'a}}, {Scholz}, and {Stairs}]{2021ApJ...908L..12T}
S.~P. {Tendulkar} et al.
\newblock \emph{\apjl}, 908\penalty0 (1):\penalty0 L12, Feb. 2021.
\newblock \doi{10.3847/2041-8213/abdb38}.

\bibitem[Thornton et~al.(2013)Thornton, Stappers, Bailes, Barsdell, Bates, Bhat, Burgay, Burke-Spolaor, Champion, Coster, D’Amico, Jameson, Johnston, Keith, Kramer, Levin, Milia, Ng, Possenti, and van Straten]{Thornton_2013}
D.~Thornton et al.
\newblock \emph{Science}, 341\penalty0 (6141):\penalty0 53–56, July 2013.
\newblock ISSN 1095-9203.
\newblock \doi{10.1126/science.1236789}.
\newblock URL \url{http://dx.doi.org/10.1126/science.1236789}.

\bibitem[{Tian} et~al.(2025){Tian}, {Pastor-Marazuela}, {Rajwade}, {Stappers}, {Shaji}, {Hanmer}, {Caleb}, {Bezuidenhout}, {Jankowski}, {Breton}, {Barr}, {Kramer}, {Groot}, {Bloemen}, {Vreeswijk}, {Pieterse}, {Woudt}, {Fender}, {Wijnands}, and {Buckley}]{Tian_2025_MNRAS}
J.~{Tian} et al.
\newblock \emph{\mnras}, 540\penalty0 (2):\penalty0 1685--1700, June 2025.
\newblock \doi{10.1093/mnras/staf793}.

\bibitem[Timmerman et~al.(2026)Timmerman, author2, author3, author4, and author5]{Timmerman01.2026.SKA}
R.~Timmerman et al.
\newblock In \emph{Advancing Astrophysics with the SKA -- II (AASKAII)}. 2026.
\newblock arXiv search: Report number AASKAII/Timmerman01.

\bibitem[{Trudu} et~al.(2023){Trudu}, {Pilia}, {Nicastro}, {Guidorzi}, {Orlandini}, {Zampieri}, {Marthi}, {Ambrosino}, {Possenti}, {Burgay}, {Casentini}, {Mereminskiy}, {Savchenko}, {Palazzi}, {Panessa}, {Ridolfi}, {Verrecchia}, {Anedda}, {Bernardi}, {Bachetti}, {Burenin}, {Burtovoi}, {Casella}, {Fiori}, {Frontera}, {Gajjar}, {Gardini}, {Ge}, {Guijarro-Rom{\'a}n}, {Ghedina}, {Hermelo}, {Jia}, {Li}, {Liao}, {Li}, {Lu}, {Lutovinov}, {Naletto}, {Ochner}, {Papitto}, {Perri}, {Pittori}, {Safonov}, {Semena}, {Strakhov}, {Tavani}, {Ursi}, {Xiong}, {Zhang}, and {Zheltoukhov}]{2023A&A...676A..17T}
M.~{Trudu} et al.
\newblock \emph{\aap}, 676:\penalty0 A17, Aug. 2023.
\newblock \doi{10.1051/0004-6361/202245303}.

\bibitem[{Van Waerbeke} and {Zhitnitsky}(2019)]{2019PhRvD..99d3535V}
L.~{Van Waerbeke} and A.~{Zhitnitsky}.
\newblock \emph{\prd}, 99\penalty0 (4):\penalty0 043535, Feb. 2019.
\newblock \doi{10.1103/PhysRevD.99.043535}.

\bibitem[{Vanderlinde} et~al.(2019){Vanderlinde}, {Liu}, {Gaensler}, {Bond}, {Hinshaw}, {Ng}, {Chiang}, {Stairs}, {Brown}, {Sievers}, {Mena}, {Smith}, {Bandura}, {Masui}, {Spekkens}, {Belostotski}, {Dobbs}, {Turok}, {Boyle}, {Rupen}, {Landecker}, {Pen}, and {Kaspi}]{2019clrp.2020...28V}
K.~{Vanderlinde} et al.
\newblock In \emph{Canadian Long Range Plan for Astronomy and Astrophysics White Papers}, volume 2020, page~28, Oct. 2019.
\newblock \doi{10.5281/zenodo.3765414}.

\bibitem[{Xu} et~al.(2022){Xu}, {Niu}, {Chen}, {Lee}, {Zhu}, {Dong}, {Zhang}, {Jiang}, {Wang}, {Xu}, {Zhang}, {Fu}, {Filippenko}, {Peng}, {Zhou}, {Zhang}, {Wang}, {Feng}, {Li}, {Brink}, {Li}, {Lu}, {Yang}, {Caballero}, {Cai}, {Chen}, {Dai}, {Djorgovski}, {Esamdin}, {Gan}, {Guhathakurta}, {Han}, {Hao}, {Huang}, {Jiang}, {Li}, {Li}, {Li}, {Li}, {Li}, {Liu}, {Luo}, {Men}, {Niu}, {Peng}, {Qian}, {Song}, {Stern}, {Stockton}, {Sun}, {Wang}, {Wang}, {Wang}, {Wang}, {Wu}, {Xiao}, {Xiong}, {Xu}, {Xu}, {Yang}, {Yang}, {Yao}, {Yi}, {Yue}, {Yu}, {Yu}, {Yuan}, {Zhang}, {Zhang}, {Zhang}, {Zhao}, {Zheng}, {Zhu}, and {Zou}]{2022Natur.609..685X}
H.~{Xu} et al.
\newblock \emph{\nat}, 609\penalty0 (7928):\penalty0 685--688, Sept. 2022.
\newblock \doi{10.1038/s41586-022-05071-8}.

\bibitem[{Zevin} et~al.(2022){Zevin}, {Nugent}, {Adhikari}, {Fong}, {Holz}, and {Kelley}]{2022ApJ...940L..18Z}
M.~{Zevin} et al.
\newblock \emph{\apjl}, 940\penalty0 (1):\penalty0 L18, Nov. 2022.
\newblock \doi{10.3847/2041-8213/ac91cd}.

\bibitem[{Zhang} et~al.(2021){Zhang}, {Zhang}, {Li}, and {Lorimer}]{2021MNRAS.501..157Z}
R.~C. {Zhang}, B.~{Zhang}, Y.~{Li}, and D.~R. {Lorimer}.
\newblock \emph{\mnras}, 501\penalty0 (1):\penalty0 157--167, Feb. 2021.
\newblock \doi{10.1093/mnras/staa3537}.

\end{thebibliography}

\end{document}